\RequirePackage{fix-cm} 
\documentclass[a4paper, twoside, reqno, dvips, 12pt]{amsart}
\usepackage{fixltx2e}   

\usepackage{etex}

\usepackage[latin1]{inputenc}
\usepackage[T1]{fontenc}

\usepackage{algorithm}
\usepackage[noend]{algpseudocode}

\usepackage[titletoc,title]{appendix}

\usepackage{esint}
\usepackage{dsfont}
\usepackage{xspace}
\usepackage{amsgen}
\usepackage{amsthm}
\usepackage{amssymb}
\usepackage{amsmath}
\usepackage{wasysym}
\usepackage{upgreek}
\usepackage{amsfonts}
\usepackage{stmaryrd}
\usepackage{mathtools}

\usepackage{relsize}
\usepackage{textcomp}
\usepackage{textgreek}
\usepackage[mathcal, mathscr]{euscript}

\usepackage{mathrsfs}
\DeclareMathAlphabet{\mathscrbf}{OMS}{mdugm}{b}{n}

\usepackage{a4wide}

\usepackage[dvipsnames, table]{xcolor}
\definecolor{bckg}{RGB}{20.8, 20.8, 20.8}
\definecolor{oneblue}{rgb}{0.0, 0.0, 0.85}
\definecolor{Lightblue}{RGB}{214, 214, 214}
\definecolor{bluepigment}{rgb}{0.2, 0.2, 0.6}
\definecolor{charcoal}{rgb}{0.21, 0.27, 0.31}
\definecolor{denimblue}{rgb}{0.08, 0.38, 0.74}
\definecolor{Lightgray}{rgb}{0.89, 0.89, 0.89}
\definecolor{darkgrey}{rgb}{0.273, 0.281, 0.30}
\definecolor{darkelectricblue}{rgb}{0.33, 0.41, 0.47}

\usepackage{multirow}

\usepackage[sort&compress, comma, square, numbers]{natbib}

\usepackage{psfrag}
\usepackage{graphicx}
\usepackage{adjustbox}
\usepackage[tight]{subfigure}
\usepackage{morefloats}
\usepackage{indentfirst}

\usepackage[usenames, dvipsnames, pdf]{pstricks}
\usepackage{epsfig}
\usepackage{pst-grad} 
\usepackage{pst-plot} 

\usepackage{rotating}
\usepackage{pdflscape}

\usepackage{acronym}
\usepackage{microtype}
\usepackage[labelsep=period,%
            labelfont={bf,sf,color=bluepigment},%
            justification=raggedright]{caption}

\usepackage[perpage, symbol]{footmisc}

\usepackage[colorlinks,
           urlcolor=oneblue,
           linkcolor=denimblue,
           citecolor=NavyBlue,
           bookmarksopen=false,
           pdfpagemode=UseNone,
           pagebackref]{hyperref}

\usepackage[explicit]{titlesec}

\titleformat{\section}[block]
  {\color{NavyBlue}\Large\sffamily\bfseries}
  {}
  {0.0em}
  {\colorbox{bckg!5}{\strut\parbox{\dimexpr\linewidth-2\fboxsep\relax}{\thesection. #1}}}
  [\vspace*{0.33em}]

\titleformat{name=\section,numberless}[block]
  {\color{NavyBlue}\Large\sffamily\bfseries}
  {}
  {0.0em}
  {\colorbox{bckg!5}{\strut\parbox{\dimexpr\linewidth-2\fboxsep\relax}{#1}}}
  [\vspace*{0.33em}]

\titleformat{\subsection}
  {\color{NavyBlue}\large\sffamily\bfseries}
  {}
  {0.0em}
  {\colorbox{bckg!5}{\parbox{\dimexpr\linewidth-2\fboxsep\relax}{\thesubsection. #1}}}
  [\vspace*{0.33em}]

\titleformat{name=\subsection,numberless}
  {\color{NavyBlue}\Large\sffamily\bfseries}
  {}
  {0em}
  {\colorbox{bckg!5}{\parbox{\dimexpr\linewidth-2\fboxsep\relax}{#1}}}
  [\vspace*{0.33em}]

\titleformat{\subsubsection}
  {\color{bluepigment}\sffamily\normalsize\bfseries}
  {\thesubsubsection}
  {0.5em}
  {#1}
  [\vspace*{0.33em}]

\titleformat{\paragraph}[runin]
  {\color{bluepigment}\sffamily\small\bfseries}
  {}
  {0em}
  {#1}

\titlespacing{\section}{0.0em}{1.5em plus 2pt minus 2pt}%
{1.0em plus 2pt minus 2pt}[0em]
\titlespacing{\subsection}{0.5em}{1.5em plus 2pt minus 2pt}%
{1.0em}[0em]
\titlespacing{\subsubsection}{0.5em}{1.5em plus 2pt minus 2pt}%
{1.0em plus 2pt minus 2pt}[0em]

\usepackage{titletoc}

\contentsmargin{0.5em}
\newlength{\tocsep} 
\titlecontents{section}[\tocsep]
  {\addvspace{10pt}\bfseries\sffamily}
  {\contentslabel[\thecontentslabel]{\tocsep}}
  {}
  {\ \titlerule*[0.75pc]{.}\ \thecontentspage}
  []
\titlecontents{subsection}[\tocsep]
  {\addvspace{8pt}\sffamily}
  {\contentslabel[\thecontentslabel]{\tocsep}}
  {}
  {\ \titlerule*[0.5pc]{.}\ \thecontentspage}
  []
\titlecontents*{subsubsection}[\tocsep]
  {\addvspace{2pt}\footnotesize\sffamily}
  {}
  {}
  {\ \titlerule*[0.35pc]{.}\ \thecontentspage}
  [\\*]

\makeatletter
\def\@setauthors{%
  \begingroup
  \def\thanks{\protect\thanks@warning}%
  \trivlist
  \centering\footnotesize \@topsep30\p@\relax
  \advance\@topsep by -\baselineskip
  \item\relax
  \author@andify\authors
  \def\\{\protect\linebreak}%
  \textsc{\normalsize\textcolor{darkelectricblue}{\authors}}%
  \ifx\@empty\contribs
  \else
    ,\penalty-3 \space \@setcontribs
    \@closetoccontribs
  \fi
  \endtrivlist
  \endgroup
}
\def\@settitle{\begin{center}%
  \baselineskip14\p@\relax
    \bfseries
    \textsc{\Large\textcolor{charcoal}{\@title}}
  \end{center}%
}
\makeatother

\usepackage{enumitem}
\setlist[description]{%
  topsep=30pt,               
  itemsep=5pt,               
  font={\bfseries\sffamily\color{NavyBlue}}, 
}

\usepackage{fancyhdr}
\usepackage{lastpage}

\newcommand*\Title{\textcolor{bluepigment}{An innovative method to determine optimum insulation thickness}}
\newcommand*\Authors{\textcolor{bluepigment}{S.~Gasparin, J.~Berger, D.~Dutykh \& N.~Mendes}}
\newcommand*{\plogo}{\textcolor{gray}{{\texttt{arXiv.org} / \textsc{hal}}}} 

\numberwithin{equation}{section}

\newcommand{\ie}{\emph{i.e.}\xspace}

\newcommand*\egal{\ = \ }
\newcommand*\plus{\ + \ }
\newcommand*\moins{\ - \ }
\newcommand*\egalb{\, = \, }

\newcommand{\cz}{c_{\,p}}
\newcommand{\czref}{c_{\,0}}
\newcommand{\hR}{h_{\,R}}
\newcommand{\hL}{h_{\,L}}

\newcommand{\ko}{k_{\,0}}

\newcommand{\Ti}{T_{\,i}}

\newcommand{\Tref}{T_{\,0}}
\newcommand{\TR}{T_{\,R}}
\newcommand{\TL}{T_{\,L}}
\newcommand{\tref}{t_{\,0}}

\newcommand{\rhoz}{\rho}

\newcommand{\Bi}{\mathrm{Bi}}
\newcommand{\BiL}{\mathrm{Bi}_{\,L}}
\newcommand{\BiR}{\mathrm{Bi}_{\,R}}
\newcommand{\Fo}{\mathrm{Fo}}
\newcommand{\cTs}{c^{\,\star}}
\newcommand{\kTs}{k^{\,\star}}
\newcommand{\qs}{q^{\,\star}}
\newcommand{\qsinf}{q^{\,\star}_{\,\infty}}

\newcommand{\ts}{t^{\,\star}}
\newcommand{\dts}{\Delta t^{\,\star}}

\newcommand{\uL}{u_{\,L}}
\newcommand{\uR}{u_{\,R}}
\newcommand{\xs}{x^{\,\star}}
\newcommand{\Rlws}{R_{\,\text{lw}}^{\,\star}}

\newcommand*\pd[2]{\dfrac{\partial #1}{\partial #2}}

\newcommand{\eqdef}{\mathop{\stackrel{\ \mathrm{def}}{:=}\ }}

\renewcommand{\L}{\mathcal{L}}
\newcommand{\abs}[1]{\lvert\, #1\, \rvert}

\newcommand{\dt}{\Delta t}
\newcommand{\dx}{\Delta x}
\renewcommand{\d}{\ensuremath{\mathrm{d}}}

\newcommand{\half}{{\frac{1}{2}}}

\newcommand{\R}{\mathds{R}}
\newcommand{\I}{\mathcal{I}}
\newcommand{\Q}{\mathcal{Q}}
\renewcommand{\P}{\mathcal{P}}
\renewcommand{\O}{\mathcal{O}\,}

\newcommand{\oC}{\mathsf{^{\circ}C}}

\begin{document}

\title[\Title]{An innovative method to determine optimum insulation thickness based on non-uniform adaptive moving grid}

\author[S.~Gasparin]{Suelen Gasparin$^*$}
\address{\textbf{S.~Gasparin:} LAMA, UMR 5127 CNRS, Universit\'e Savoie Mont Blanc, Campus Scientifique, F-73376 Le Bourget-du-Lac Cedex, France and Thermal Systems Laboratory, Mechanical Engineering Graduate Program, Pontifical Catholic University of Paran\'a, Rua Imaculada Concei\c{c}\~{a}o, 1155, CEP: 80215-901, Curitiba -- Paran\'a, Brazil}
\email{suelengasparin@hotmail.com}
\urladdr{https://www.researchgate.net/profile/Suelen\_Gasparin/}
\thanks{$^*$ Corresponding author}

\author[J.~Berger]{Julien Berger}
\address{\textbf{J.~Berger:} LOCIE, UMR 5271 CNRS, Universit\'e Savoie Mont Blanc, Campus Scientifique, F-73376 Le Bourget-du-Lac Cedex, France}
\email{Berger.Julien@univ-smb.fr}
\urladdr{https://www.researchgate.net/profile/Julien\_Berger3/}

\author[D.~Dutykh]{Denys Dutykh}
\address{\textbf{D.~Dutykh:} Univ. Grenoble Alpes, Univ. Savoie Mont Blanc, CNRS, LAMA, 73000 Chamb\'ery, France and LAMA, UMR 5127 CNRS, Universit\'e Savoie Mont Blanc, Campus Scientifique, F-73376 Le Bourget-du-Lac Cedex, France}
\email{Denys.Dutykh@univ-smb.fr}
\urladdr{http://www.denys-dutykh.com/}

\author[N.~Mendes]{Nathan Mendes}
\address{\textbf{N.~Mendes:} Thermal Systems Laboratory, Mechanical Engineering Graduate Program, Pontifical Catholic University of Paran\'a, Rua Imaculada Concei\c{c}\~{a}o, 1155, CEP: 80215-901, Curitiba -- Paran\'a, Brazil}
\email{Nathan.Mendes@pucpr.edu.br}
\urladdr{https://www.researchgate.net/profile/Nathan\_Mendes/}


\begin{titlepage}
\thispagestyle{empty} 
\noindent
{\Large Suelen \textsc{Gasparin}}\\
{\it\textcolor{gray}{Pontifical Catholic University of Paran\'a, Brazil}}\\
{\it\textcolor{gray}{LAMA--CNRS, Universit\'e Savoie Mont Blanc, France}}
\\[0.02\textheight]
{\Large Julien \textsc{Berger}}\\
{\it\textcolor{gray}{LOCIE--CNRS, Universit\'e Savoie Mont Blanc, France}}
\\[0.02\textheight]
{\Large Denys \textsc{Dutykh}}\\
{\it\textcolor{gray}{LAMA--CNRS, Universit\'e Savoie Mont Blanc, France}}
\\[0.02\textheight]
{\Large Nathan \textsc{Mendes}}\\
{\it\textcolor{gray}{Pontifical Catholic University of Paran\'a, Brazil}}
\\[0.10\textheight]

\colorbox{Lightblue}{
  \parbox[t]{1.0\textwidth}{
    \centering\huge\sc
    \vspace*{0.7cm}
    
    \textcolor{bluepigment}{An innovative method to determine optimum insulation thickness based on non-uniform adaptive moving grid}

    \vspace*{0.7cm}
  }
}

\vfill 

\raggedleft     
{\large \plogo} 
\end{titlepage}


\newpage
\thispagestyle{empty} 
\par\vspace*{\fill}   
\begin{flushright} 
{\textcolor{denimblue}{\textsc{Last modified:}} \today}
\end{flushright}


\newpage
\maketitle
\thispagestyle{empty}


\begin{abstract}

It is well known that thermal insulation is a leading strategy for reducing energy consumption associated to heating or cooling processes in buildings. Nevertheless, building insulation can generate high expenditures so that the selection of an optimum insulation thickness requires a detailed energy simulation as well as an economic analysis. In this way, the present study proposes an innovative non-uniform adaptive method to determine the optimal insulation thickness of external walls. First, the method is compared with a reference solution to properly understand the features of the method, which can provide high accuracy with less spatial nodes. Then, the adaptive method is used to simulate the transient heat conduction through the building envelope of buildings located in \textsc{Brazil}, where there is a large potential of energy reduction. Simulations have been efficiently carried out for different wall and roof configurations, showing that the innovative method efficiently provides a gain of $25\%$ on the computer run time.


\bigskip\bigskip
\noindent \textbf{\keywordsname:} Optimum insulation thickness; thermal insulation; numerical simulation; moving grid method; redistribution schemes; adaptive numerical methods \\


\end{abstract}


\newpage
\tableofcontents
\thispagestyle{empty}


\newpage
\section{Introduction}

In \textsc{Brazil}, the electricity consumption in residential, commercial and public sectors accounted for $42.8\%$ of the \textsc{Brazilian} electricity consumption in $2016$ \cite{Filho2017}. The consumption of electrical energy in buildings is related to gains or losses of heat through the building envelope that is also associated with the internal loads. In particular, the use of heating, ventilation, and air conditioning (HVAC) systems can represent $40\%$ of the total electrical energy consumption \cite{ABRAVA2019}.

Due to current environmental issues, there is a great potential to reduce heating and cooling loads particularly through improvements of thermal performance of the envelope. One of the most efficient ways to reduce the transmission loads and, consequently, reduce the energy consumption is the use of an appropriated thermal insulation for the building envelope. However, in \textsc{Brazil}, the insulation costs are relatively high, due especially to the low demand and the lack of a compulsory regulation \cite{Morishita2016}. Moreover, the choice of the insulation thickness is a difficult task since it is a delicate compromise between the cost of insulation materials, which increases with the thickness, and the energy cost for heating/cooling the building, that decreases with the thickness \cite{Daouas2011, KameniNematchoua2017}.

The determination of the optimum thickness is a difficult task. It requires the computation of the heating or cooling loads of a building along a whole year by solving the heat transfer equation through building walls and roofs. Several studies reported in literature enhance the importance of efficient modeling to assess wall thermal loads. The influence of important wall parameters on the energy efficiency has been investigated: insulation properties in \cite{Ioannou2015}, thermal inertia in \cite{Aste2009, Taylor2014} or thermal conductivity in \cite{Kontoleon2013}. In \cite{Kontoleon2016}, the effect of wall moisture content on the thermal response of the wall has been analysed. Some studies aimed at calibrating the results of the numerical models with observations in order to improve the reliability of the physical model, as for instance in \cite{Mustafaraj2014, Kim2017, Yang2015, Yuan2017}.

These studies highlight the importance of efficient numerical models to predict heat transfer through walls. However, the elaboration of such models is a difficult issue since the numerical complexity arises from several points \cite{Mendes2017, Capizzi2017}. First, the numerical model has to deal with different time scales since the computation is carried out over one year, as illustrated in \cite{Branco2004}, while the characteristic time scale of the heat diffusion scales with $15 \ \mathsf{min}\,$ or even less. In addition, the boundary conditions are defined according to non-periodic climatic data \cite{Reilly2017}. Third, the wall configuration involves several layers inducing variations of the thermal properties with space. These difficulties imply an important numerical effort of the numerical model to compute the optimal thermal insulation.

In literature, several works circumvent the numerical difficulties to reduce the computational efforts. For instance, in \cite{Capizzi2017}, a whole building model is replaced by lumped approaches, denoted as RC model to reduced the computation efforts. This approach is used in many algorithms to solve the partial differential equation of heat transfer as mentioned in \cite{Fraisse2002, Roels2017, Naveros2015}. For the study of optimal thermal insulation, studies also attempt to avoid the numerical difficulties. For instance, in \cite{Ekici2012}, the simple degree-day method is used to compute the building energy consumption in four cities of \textsc{Turkey}. In \cite{KameniNematchoua2017, Daouas2011, Axaopoulos2014} simulations are performed only on the hottest and coldest days/months of the year. In \cite{Ozel2011}, the outside boundary condition is assumed as periodic by repeating a daily cycle of air temperature and solar radiation. In \cite{Al-Sanea2005, Al-Sanea2012}, a representative day for each month of the year is used to compute the yearly loads. In \cite{Daouas2016}, an analytical solution is proposed for simulations carried out only for the hottest and coldest days of the year. For the climate of \textsc{Paris}, \cite{Berger2017d} proposed a parametric solution of the transient heat transfer using a reduced order model to determine the annual loads and consequently the optimum insulation thickness.

In this way, we propose an innovative method to simulate the transient heat transfer to determine the optimum insulation thickness, which will be tested in \textsc{Brazilian} buildings context. Instead simulating the hottest and coolest months we simulate the whole year to get the transmission loads. The method used for simulations is called Quasi-Uniform Nonlinear Transformation (QUNT) which is adaptive regarding the spatial grid \cite{Khakimzyanov2015a}. The grid points of the spatial mesh move following the monitor function which indicates their placement. The adapted grid is obtained according to the gradients of the problem indicated by solving a parabolic problem. As shown in \cite{Khakimzyanov2015b}, this approach upgrades the usual finite-differences method, with a semi-implicit scheme, to the fourth order by using the non-uniform grid \cite{Blom1994}. Just by changing the distribution of nodes one can gain two extra orders of accuracy, which allows simulating with efficiency using less discrete points. This numerical model enables to compute the temperature distribution for the entire year at a low computational effort. From these results, the optimal thermal insulation can be easily computed.

Therefore, this paper first describes in Section~\ref{sec:math_model} the employed physical model of heat transfer. In Section~\ref{sec:numerical_model}, basics of the moving grid method is detailed before exploring the features of the scheme applied to a nonlinear case, presented in Section~\ref{sec:benchmark}. Then, in Section~\ref{sec:application}, the method is applied to investigate the optimum insulation thickness in \textsc{Brazilian} buildings. The main conclusions of this study are outlined in Section~\ref{sec:conclusion}.


\section{Mathematical formulation of the physical problem}
\label{sec:math_model}

The physical problem considers one-dimensional heat transfer through a porous material. The spatial domain is defined by $x\, \in\, [\, 0, \, l \,] $, in which $x \egal 0$ corresponds to the surface in contact with the inside room and, $x \egal l$, corresponds to the outside surface. The heat transfer is governed by conduction, which formulation is derived from \textsc{Fourier}'s law:
\begin{align}\label{eq:heat_equation}
  & c \ \pd{T}{t} \egal \pd{}{x} \Biggl(\, k \ \pd{T}{x} \,\Biggr) \,,
\end{align}
where $x$ is the spatial coordinate $[\mathsf{m}]$, $t$ is the time $[\mathsf{s}]$, $T$ is the temperature $[\mathsf{K}]$\footnote{The results are presented in $[\oC]$ but simulations were performed considering $[\mathsf{K}]$. }, $k$, the thermal conductivity $[\mathsf{W/(m\cdot K)}]$ and $c \egalb \rhoz \, \cz\,$, the thermal storage, in which $\cz$ is the material specific heat $[\mathsf{J/(kg\cdot K)}]$ and $\rhoz$, the material density $[\mathsf{kg/m^3}]$.

The \textsc{Dirichlet}-type boundary conditions are written as:
\begin{align*}
  T\,(x\,=\,0\,,t) \egal \TL\, (t)  & & \text{and}  & & T\,(x\,=\,l\,,t) \egal \TR\, (t) \,.
\end{align*}
Moreover, the internal \textsc{Robin}-type boundary condition is expressed as:
\begin{align*}
  -\,k \, \pd{T}{x} \biggr\vert_{x \egalb l} &\egal \hR \, \Bigl(\, T \moins \TR \, (t)\,\Bigr)  \,,
\end{align*}
and on the external side, the boundary condition includes the convection, short- and long-wave radiation:
\begin{align*}
  k \, \pd{T}{x} \biggr\vert_{x \egalb 0} &\egal \hL \, \Bigl(\, T \moins \TL\, (t) \,\Bigr) \moins \alpha\, q_{\,\infty}\, (t) \plus \epsilon\, \sigma\, \Bigl(\, T^{\,4} \moins T_{\,\text{sky}}^{\,4}\, (t) \,\Bigr)  \,.
\end{align*}
where $\TL$ and $\TR$ are the temperature of the air $[\mathsf{K}]$ that varies over time, $h$ is the convective heat transfer coefficient $[\mathsf{W/(m^2\cdot K)}]$ and subscripts $L$ and $R$ represent the left and right boundaries. If the bounding surface is in contact with the outside building air, $q_{\,\infty}$ is the total solar radiation $[\mathsf{W/m^2}]\,$, which includes the direct, diffuse and reflexive radiations and $\alpha$ is the surface absorptivity $[-]$. For the long-wave radiation, $T_{\,\text{sky}}$ is the sky temperature $[\mathsf{K}]$, $\epsilon$ is the emissivity $[-]$ and $\sigma$ is the \textsc{Stefan--Boltzmann} constant $[\mathsf{W/(m^2\cdot K^4)}]\,$.

For the initial condition, a linear temperature distribution in function of $x$ is considered:
\begin{align}\label{eq:ic}
  T\, (x\,,t \,=\, 0)  &\egal \Ti\, (x)\,.
\end{align}

The heat flux density $q\ [\mathsf{W/m^2}]$ at the point $x_{\,j}$ is computed as:
\begin{align}\label{eq:flux}
  q_{\,j}\,(t) \eqdef -\, k\, \pd{T}{x}\,(x\,,t) \biggr\vert_{\,x \egalb x_{\,j}} \,.
\end{align}

Finally, the integration of the resulting heat flux density during a time period will give the transmission loads $E\ [\mathsf{J/m^2}]$  \cite{Axaopoulos2014, Berger2017d}:
\begin{align}\label{eq:load}
  E \egal \int_{t_0}^{t_1} q\, (\tau)\, \d \tau \,,
\end{align}
which can be evaluated on daily or monthly periods.


\section{Method of solution}
\label{sec:numerical_model}

The problem previously described is solved by using the Quasi-Uniform Nonlinear Transformation (QUNT) \cite{Khakimzyanov2015a}. However, the problem must be expressed in a dimensionless formulation before being solved numerically. For example, this process can reduce the number of parameters of the problem allowing direct comparisons of models. For a detailed explanation of why this practice is important one can consult \cite{Gasparin2017}.

In this way, we define the following dimensionless quantities:
\begin{align*}
  u \ &\eqdef \ \frac{T}{\Tref} \,, 
  &\xs \ &\eqdef \ \frac{x}{l} \,, 
  &\ts \ &\eqdef \ \frac{t}{\tref} \,,  \\[3pt]
  \cTs \ &\eqdef \ \frac{c }{\czref} \,,
  &\kTs \ &\eqdef \ \frac{k }{\ko} \,, 
  &\Bi \ &\eqdef \ \frac{h \cdot l}{\ko} \,, \\[3pt]
  \Fo \ &\eqdef \ \frac{\tref \cdot \ko}{l^{\,2} \cdot \czref} \,, 
  &\qs \ &\eqdef \ \frac{l\cdot q}{\Tref \cdot \ko}  \,,
  &\Rlws  \ &\eqdef \ \frac{\epsilon \cdot \sigma \cdot l \cdot \Tref^{\,3}}{\ko} \,,
\end{align*}
where the subscript $0$ represents a reference value, chosen according to the application problem and the superscript $\star$ represents a dimensionless quantity of the same variable. Therefore, the governing Equation~\eqref{eq:heat_equation} can be written in a dimensionless form as:
\begin{align}\label{eq:heat_equation_dimless}
  \cTs \ \pd{u}{\ts} & \egal \Fo \, \pd{}{\xs} \biggl(\, \kTs \ \pd{u}{\xs} \, \biggr)\,,  
\end{align}
with the following dimensionless formulation of the \textsc{Dirichlet} and \textsc{Robin}-type boundary conditions:
\begin{align*}
  u\,(\xs\,=\,0\,,\ts) &\egal \uL\, (\ts) \,, \\
  u\,(\xs\,=\,1\,,\ts) &\egal \uR\, (\ts)  \,, \\
  \kTs \ \pd{u}{\xs} \biggr\vert_{\xs \egalb 0} &\egal \BiL \, \Bigl(\, u \moins \uL\, (\ts) \,\Bigr) \moins \alpha\, \qsinf\, (\ts) \plus \Rlws \,\Bigl(\, u^{\,4} \moins u^{\,4}_{\,\text{sky}} \, (\ts) \,\Bigr) \,, \\
  -\ \kTs \ \pd{u}{\xs} \biggr\vert_{\xs \egalb 1} &\egal \BiR \, \Bigl(\, u \moins \uR\,(\ts)\,\Bigr) \,.
\end{align*}
For the initial condition the dimensionless form is:
\begin{align*}
  u\,(\xs\,,\ts\,=\,0) &\egal u_{\,i}\,(\xs) \,.
\end{align*}
Moreover, the dimensionless form of the heat flux density is written as:
\begin{align*}
  \qs_{\,j} \, (\ts) &\egal -\, \kTs \,\pd{u}{\xs} \biggr\vert_{\,\xs\,=\,\xs_{\,j}} \,.
\end{align*}

In the following, we drop $\star$ for the sake of notation compactness but without losing generality.


\subsection{Quasi-Uniform Nonlinear Transformation}

The Quasi-Uniform Nonlinear Transformation (QUNT) \cite{Khakimzyanov2015a, Khakimzyanov2015b, Blom1994} is a numerical method with an adaptive spatial grid which provides a high resolution where it is required. The adaptivity is obtained by using a monitor function which indicate the location of the spatial grid points during the simulation. The adapted grid is obtained by means of the equidistribution principle. This moving grid nodes approach has the advantage of being conservative in space and also second-order accurate.

Before solving the problem on a moving mesh, one have to chose a robust and accurate scheme on a fixed grid. This scheme will be then generalized to incorporate the motion of the mesh. This scheme is described in the sequence.


\subsubsection{A finite-difference scheme on a fixed uniform mesh}

Consider the dimensionless form of the heat diffusion Equation~\eqref{eq:heat_equation_dimless} with a simplified notation, just for describing the numerical method:
\begin{align}\label{eq:diff_admen_simpl}
  c\,(u) \ \pd{u}{t} & \egal \pd{}{x} \biggl(\, k\,(u) \ \pd{u}{x} \biggr)\,. 
\end{align}
The spatial domain $\I \egalb [\,0\,,l\,]$ is uniformly discretized, with $x_{\,j} \egalb j \dx,\, j \in \mathbb{Z}$ being the spatial nodes, in which $\dx \, >\, 0$ is the spatial step size. For the time domain $\P \egalb [\,0,\,\tau\,]$, also consider a uniform discretization, where $t^{\,n} \egalb n\, \dt\,, n \in \mathbb{Z}^{+}$ are the temporal nodes and $\dt\, >\, 0$ being the time step. 
Thus, by using an IMplicit-EXplicit (IMEX) scheme, Equation~\eqref{eq:diff_admen_simpl} is written in a discrete form:
\begin{align*}
  c_{\,j}^{\,n}\, \dfrac{u_{\,j}^{\,n+1} \moins u_{\,j}^{\,n}}{\dt} \egal \dfrac{1}{\dx}\, \Biggl[k_{\,j+\half}^{\,n} \, \dfrac{u_{\,j+1}^{\,n+1} \moins u_{\,j}^{\,n+1}}{\dx }  \moins k_{\,j-\half}^{\,n} \, \dfrac{u_{\,j}^{\,n+1} \moins u_{\,j-1}^{\,n+1}}{\dx}\Biggr] \,.
\end{align*}
This scheme approximates the continuous operator to order $\O\,(\dx^{\,2}\ +\ \dt)$. The advantage of this semi-implicit scheme over the fully implicit is to avoid sub-iterations in the solution procedure and, at the same time, being stable and consistent. For simplicity, the boundary conditions are not treated here. In what follows, the method based on this scheme with the moving grid is describe.


\subsubsection{Finite-difference scheme on a moving grid}

Consider the computational domain $\I \egalb [\,0\,,l\,]\,$, the reference domain $\Q \egalb [\,0\,,1\,] $ and a bijective time-depend mapping from $\Q$ to $\I\,$:
\begin{align*}
  x\,(q,\, t): \Q\times \R^{\,+}\ \longmapsto\ \I \,.
\end{align*}
This represents the quasi-uniform nonlinear transformation. In addition, the boundary points are required to map into each other:
\begin{align*}
  x\,(0\,,t) \egal 0 & &\text{and} & & \quad \quad x\,(1\,,t) \egal l\,.
\end{align*}
The reference domain $\Q$ is uniformly discretized into $N$ elements, with $q_{\,j}\egalb j\,h,\, j \in \mathbb{Z}^{+}$ representing the nodes of the grid, which is equally spaced $h\egalb 1/N\,$. However, only the image of nodes $q_{\,j}$ under the map $x\,(q\,,t)$ are needed, since they constitute the nodes of the moving mesh:
\begin{align*}
  x\, (q_{\,j},\,t^{\,n}) \egal x_{\,j}^{\,n} \,.
\end{align*}
The diffusion Equation~\eqref{eq:diff_admen_simpl} is rewritten on the domain $\Q$, with the help of the composed function $v\,(q\,,t) \eqdef (u \circ x)\,(q\,,t) \, \equiv\, u\, \Bigl(x\,(q\,,t)\,,t \Bigr)\,$:
\begin{align*}
  J\,(q\,,t) \, c\,(v)\, \pd{v}{t} \egal \pd{}{q} \Bigg[\dfrac{k(v)}{J\,(q\,,t)}\, \pd{v}{q}\Bigg] \plus c\,(v)\, \pd{v}{q}\, \pd{x}{t}\,,
\end{align*}
where $J \eqdef \pd{x}{q}$ is the \textsc{Jacobian} of the transformation $x\,(q\,,t)\,$.

Therefore, the discrete form of the diffusion heat equation on a moving mesh grid is:
\begin{align*}
  J_{\,j}^{\,n}\, c_{\,j}^{\,n}\, \dfrac{v_{\,j}^{\,n+1} \moins v_{\,j}^{\,n}}{\dt} \egal & \dfrac{1}{h}\, \Biggl[\,\dfrac{k_{\,j+\half}^{\,n}}{J_{\,j+\half}^{\,n}}\, \Bigg(\dfrac{v_{\,j+1}^{\,n+1} \moins v_{\,j}^{\,n+1}}{h}\Bigg) \moins \dfrac{k_{\,j-\half}^{\,n}}{J_{\,j-\half}^{\,n}}\, \Bigg(\dfrac{v_{\,j}^{\,n+1} \moins v_{\,j-1}^{\,n+1}}{h}\Bigg)   \Biggr]\plus  \nonumber \\
  & \plus c_{\,j}^{\,n} \, \Bigg(\dfrac{v_{\,j+1}^{\,n+1} \moins v_{\,j-1}^{\,n+1}}{2\,h}\Bigg)\cdot\Bigg(\dfrac{x_{\,j}^{\,n+1} \moins x_{\,j}^{\,n}}{\dt }\Bigg)  \,.
\end{align*}
In summary, this is the parametrization process of the mesh motion by a bijective mapping. In the sequence, the construction of this map is explained.


\subsubsection{Construction and motion of the grid}

The equidistribution method is used to construct the adaptive grid. To control the distribution of the nodes, a positive valued function $w\, (x,\,t)$ has to be chosen. This function is called \textit{monitor function}. The choice of the monitor function and its parameters are important for the accuracy of the scheme. In particular, for unsteady nonlinear problems, there are several choice options. One can find more information, for example in \citep{Stockie2001}. In this work, the function chosen is based on \cite{Khakimzyanov2015a} and can be written as:
\begin{align}\label{eq:monitor_function}
  w\,(x,t) \egal 1 \plus \alpha_{\,1} \, \abs{u}^{\,\beta_{\,1}} \plus \alpha_{\,2}\, \abs{\pd{u}{x}}^{\,\beta_{\,2}}\,,
\end{align}
where $\alpha_{\,1}\,$, $\alpha_{\,2}\,$, $\beta_{\,1}$ and $\beta_{\,2}$ are positive real parameters. The quantity $w\,(x\,,t)$ assumes large positive values in areas where the solution has important gradients, inducing the elements of the grid mesh to approach. The discrete formulation of the monitor function presented in Eq.~\eqref{eq:monitor_function} is written as:
\begin{align*}
  w_{\,j}^{\,n} \egal 1 \plus \alpha_{\,1}\, \abs{u_{\,j}^{\,n}}^{\, \beta_1} \plus \alpha_{\,2}\, \abs{\dfrac{u_{\,j+1}^{\,n} \moins u_{\,j}^{\,n}}{x_{\,j+1}^{\,n} \moins x_{\,j}^{\,n}}}^{\,\beta_2} \,.
\end{align*}

A non-uniform grid $\I$ is given if we construct the mapping $x\,(q\,,t)\,:\ \Q\ \longmapsto\ \I$ and evaluate it in the nodes of the uniform grid. To this end, the equidistribution method proposes that the desired mapping $x\,(q\,,t)$ be obtained as a solution of a nonlinear parabolic problem, which is written as:
\begin{align}\label{eq:parabolic_problem}
  \pd{}{q}\biggl(w\, (x\,,t)\, \pd{x}{q} \biggr) \egal \beta \, \pd{x}{t}\,,
\end{align}
with $x\,(0,\,t) \egal 0$ and $x\,(1,\,t) \egal l$ as the boundary conditions. Eq.~\eqref{eq:parabolic_problem} can be written in the discrete form as:
\begin{align*}
  \dfrac{1}{h}\ \Biggl( w^{\, n}_{j+\half}\dfrac{x^{n+1}_{j+1} \moins x^{n+1}_{j}}{h} \moins w^{\, n}_{j-\half} \dfrac{x^{n+1}_{j+1} \moins x^{n+1}_{j}}{h}\Biggr) \egal \beta\ \dfrac{x^{n+1}_{j} \moins x^{n}_{j}}{\dt} \,,
\end{align*}
with the same boundary conditions $x_{\,0}^{\,n\,+\,1} \egalb 0$ and $x_{\,N}^{\,n\,+\,1} \egal l\,$. The parameter $\beta \ >\ 0$ plays the role of the inverse diffusion coefficient and it controls the smoothness of nodes trajectory.

\bigskip
\paragraph{Initial grid generation.}

As the problem depends also on time, it is of capital importance to obtain a high-quality initial mesh. The initial condition must be adapted to the new grid before starting the dynamical simulation. At $t \egalb 0$ we compute the monitor function $w\,(x,\,0)$ of the initial condition $u\,(x,\,0)\,$. Then, the mapping $x\,(q,\,0)$ is determined as the solution of Eq.~\eqref{eq:parabolic_problem}, which is reduced to a simple second-order ordinary differential equation:
\begin{align*}
  \pd{}{q}\biggl(w\, (x\,,0)\, \pd{x}{q} \biggr) \egal 0 \,,
\end{align*}
supplemented with the \textsc{Dirichlet} boundary condition. Thus, the finite-difference approximation of this latter differential equation is:
\begin{align*}
  \dfrac{1}{h}\ \Biggl( w^{\, 0}_{\,j+\half}\dfrac{x^{\,0}_{\,j+1} \moins x^{\,0}_{\,j}}{h} \moins w^{\,0}_{\,j-\half} \dfrac{x^{\,0}_{\,j+1} \moins x^{\,0}_{\,j}}{h} \, \Biggr) \egal 0 \,,
\end{align*}
with the discrete boundary conditions $x_{\, 0}^{\, 0} \egalb 0\,$, $x^{\, 0}_{\, N} \egalb l\,$. This latter nonlinear system of equations is solved iteratively, with iterations initialized with a uniform grid as the first guess. Its solution satisfies the equidistribution principle: in areas where $w^{\, 0}_{\,j\,+\,\half}$ takes large values, the space between two neighboring nodes $x^{\,0}_{\,j+1}$ and $x^{\,0}_{\,j}$ has to be inversely proportionally small. One can find more information in \cite{Khakimzyanov2015b}.

\bigskip
\paragraph{Smoothing step.}

To ensure the smoothness of the mesh motion the monitor function $w$ can be filtered. This process, called \textit{smoothing step} enable to enlarge the set of acceptable values of the parameters $\alpha$ and $\beta$ of the monitor function. According to \cite{Khakimzyanov2015a}, the following implicit scheme can produce robust results:
\begin{align}
  \bar{w}_{\,j\,+\,\half} \egal w_{\,j\,+\,\half} \moins \sigma \, \bar{w}_{\,j\,+\,\half} \plus \dfrac{\sigma}{2}\, \Bigl( \bar{w}_{\,j\,-\,\half} \plus \bar{w}_{\,j\,+\,\frac{3}{2}}\Bigr)\,,\label{eq:smooth_sys}
\end{align}
where $\sigma$ is a positive smoothing parameter. To complete Eq.~\eqref{eq:smooth_sys}, boundary conditions are taken as in the original problem:
\begin{align*}
  & \bar{w}_{\,\half} \egal w_{\,\half}\,,\quad \quad \bar{w}_{\,N\,-\,\half} \egal w_{\,N\,-\,\half}\,.
\end{align*}
The smoothed discrete monitor function $\Bigl\{\bar{w}_{\,j\,+\,\half}\Bigr\}_{\,j\,=\,0}^{\,N\,-\,1}$ is then obtained by solving the linear System~\eqref{eq:smooth_sys}. Note that the smoothing operator is applied to the monitor function.

\bigskip
\paragraph{On the choice of the parameters.}

The parameters $\alpha_{\,1}\,$, $\alpha_{\,2}\,$, $\beta_{\,1}\,$, $\beta_{\,2}\,$, $\beta$ and $\sigma$ are chosen according to the problem and the solution under consideration. To determine the optimal values one has to make some numerical experiments. Interested readers are invited to consult \cite{Khakimzyanov2015a} for more details on this aspect.


\subsection{Validation of the numerical solution}
\label{sec:validat}

All the numerical results in this paper were computed using \texttt{Matlab\texttrademark} \cite{Matlab}. The reference solution is computed using the open source package \texttt{Chebfun} \citep{Driscoll2014}. Using \texttt{pde15s} function, it enables to compute a numerical solution of a partial derivative equation with the \textsc{Chebyshev} polynomials adaptive spectral methods.

The \textsc{Euclidean} error (distance) for a determined field $u$ at time $t^{\,n}$ is computed as: 
\begin{align*}
  \delta\,(x)\ \eqdef\ \sqrt{\,\Bigl(u^{\,\text{ref}}\,(x\,,t^{\,n}) \moins u^{\,\text{num}}\,(x\,,t^{\,n}) \Bigr)^{\,2}} \,,
\end{align*}
where $u^{\,\text{ref}}$ is the reference solution computed with \texttt{Chebfun} and $u^{\,\text{num}}$ is the solution computed with the QUNT or classical schemes.

The error over the time is computed as a function of $x$ by the following formulation:
\begin{align*}
  \varepsilon\, (x)\ &\eqdef\ \sqrt{\,\frac{1}{N_{\,t}} \, \sum_{n\, =\, 1}^{N_{\,t}} \, \Bigl( \, u^{\, \text{ref}}\, (x\,, t^{\, n}) \moins u^{\,\text{num}}\, (x\,, t^{\, n}) \, \Bigr)^{\,2}}\,,
\end{align*}
where $N_{\,t}$ is the number of temporal steps. The global $\L_{\,\infty} $ error is given by the maximum value of $\varepsilon\, (x)\,$: 
\begin{align*}
  \varepsilon_{\, \infty}\ &\eqdef\ \sup_{x \ \in \ [\, 0 \,,\, l \,]} \, \varepsilon\, (x) \,.
\end{align*}

For the heat flux density, the \textsc{Euclidean} error (distance) is computed as:
\begin{align*}
  \xi\, (t) \eqdef \sqrt{\Bigl(q^{\,\text{ref}}\, (t)  \moins q^{\,\text{num}}\, (t) \Bigr)^2 }\,,
\end{align*}
where $q^{\,\text{ref}}$ is the reference flux density computed with \texttt{Chebfun} solution and $q^{\,\text{num}}$ is the flux density computed with the QUNT or classical solutions. The global $\L_{\,\infty} $ error of the flux density is also given by the maximum value but of the $\xi\,(t)\,$:
\begin{align*}
  \xi_{\, \infty}\ &\eqdef\ \sup_{t \ \in \ [\, 0 \,,\, \tau \,]} \, \xi\, (t) \,.
\end{align*}


\section{Numerical Benchmarking }
\label{sec:benchmark}

\begin{figure}
\centering
\includegraphics[width=0.5\textwidth]{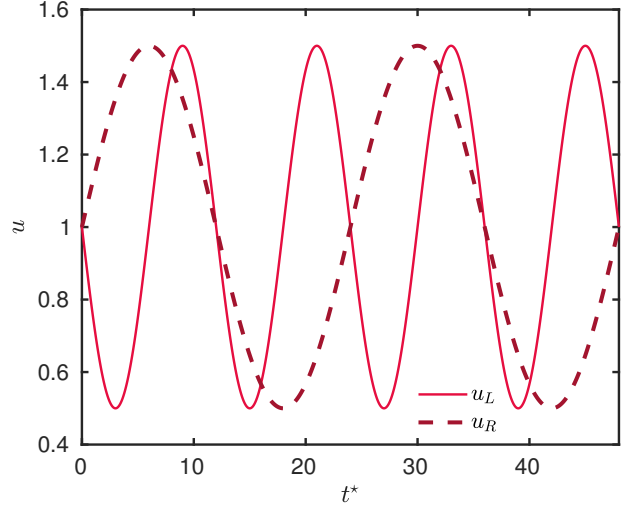}
\caption{\small\em Dimensionless ambient temperatures.}
\label{Figure:AN1_boundary}
\end{figure}

The purpose of this section is to validate the QUNT method by comparing it with the IMEX scheme and the reference solution. The problem is inspired by \cite{Gasparin2017}. As the purpose is to evaluate the efficiency of the QUNT approach, and not analyzing the physical phenomena, the study is performed using the dimensionless problem described by Equation~\eqref{eq:heat_equation_dimless} with \textsc{Dirichilet}-type boundary conditions and the following dimensionless material properties:
\begin{align*}
  & \kTs\,(u) \egal 1 \plus 0.91 \, u \plus 600 \cdot \exp \Bigl[ \,-10 \, \bigl(\, u \moins 1.5 \, \bigr)^{\,2} \, \Bigr] \,,\\
  & \cTs\,(u) \egal 900 \moins 656 \, u \plus 10^4 \cdot \exp \Bigl[ \,-5 \, \bigl(\, u \moins 1.5 \, \bigr)^{\,2} \,  \Bigr] \,.
\end{align*}
The dimensionless ambient temperature at the left and right boundaries are:
\begin{align*}
  \uL \,(\ts)  \egal 1 \moins 0.5 \cdot \sin\, \biggl( \dfrac{2\,\pi\, \ts}{12} \biggr)  & & \text{and} & & \uR \,(\ts)  \egal 1 \plus 0.5 \cdot \sin\, \biggl( \dfrac{2\,\pi\, \ts}{24} \biggr)\,,
\end{align*}
which are graphically given in Figure~\ref{Figure:AN1_boundary}. The initial temperature is considered uniform over the material and its dimensionless formulation is given by:
\begin{align*}
  u \,(\,x,\, 0\,)  \egal 1 \,.
\end{align*}

The monitoring function is defined by Equation~\eqref{eq:monitor_function} with following values for its parameters: 
\begin{align*}
  \alpha_{\,1} \egal 0.9\,, & & \beta_{\,1} \egal 2\,, & & \alpha_{\,2} \egal 0.1\,, & & \beta_{\,2} \egal 2\,.
\end{align*}
These parameters were determined by our numerical investigations and correspond to the best combination for this case study. For the grid diffusion parameter, the best value tested was $\beta \ =\ 100$ and for the smoothing parameter $\sigma \ =\ 10\,$. Simulations are performed up to the final time of $\tau \ =\ 48\,$, with $\dts \ =\ 5\cdot 10^{\,-\,3}$ and a number of spatial points $N_{\,x} \ =\ 51\,$, at first fixed for both methods.

\subsection{Results and discussion}

\begin{figure}
\centering
\includegraphics[width=0.85\textwidth]{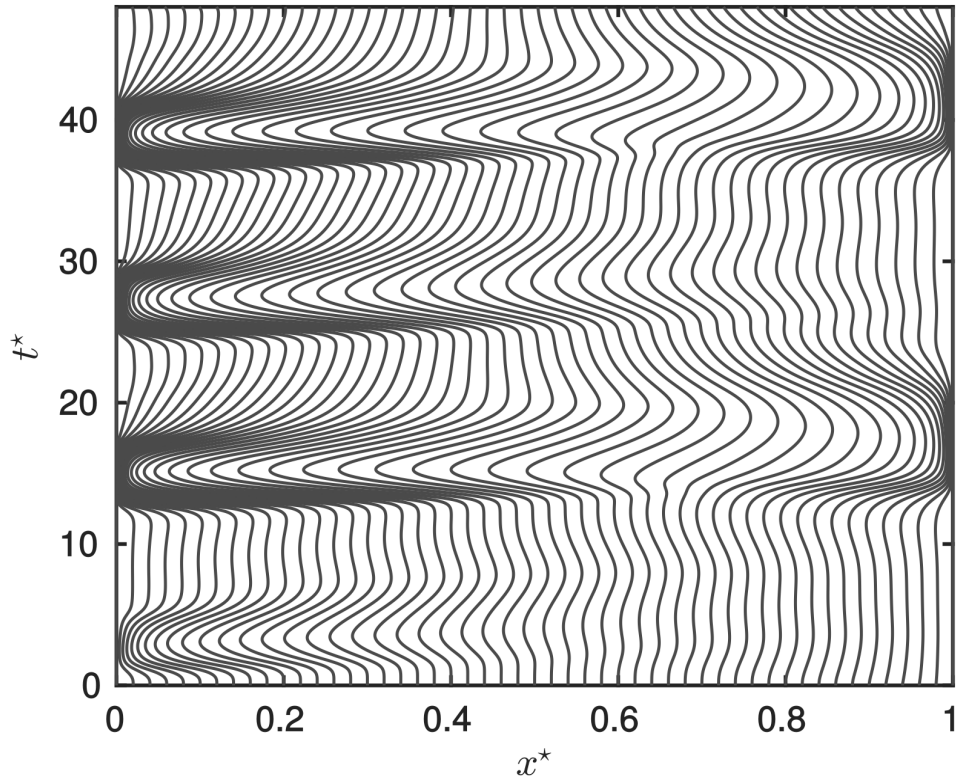}
\caption{\small\em Trajectory of the grid nodes in space-time.}
\label{Figure:AN1_tragectory}
\end{figure}

The sample trajectory of the spatial nodes can be observed in Figure~\ref{Figure:AN1_tragectory}. The spatial points concentrate where the values of the gradient are high, particularly on the boundaries where important variations of the fields are imposed.

\begin{figure}
\centering
\subfigure[a][\label{Figure:AN1_profile}]{\includegraphics[width=0.48\textwidth]{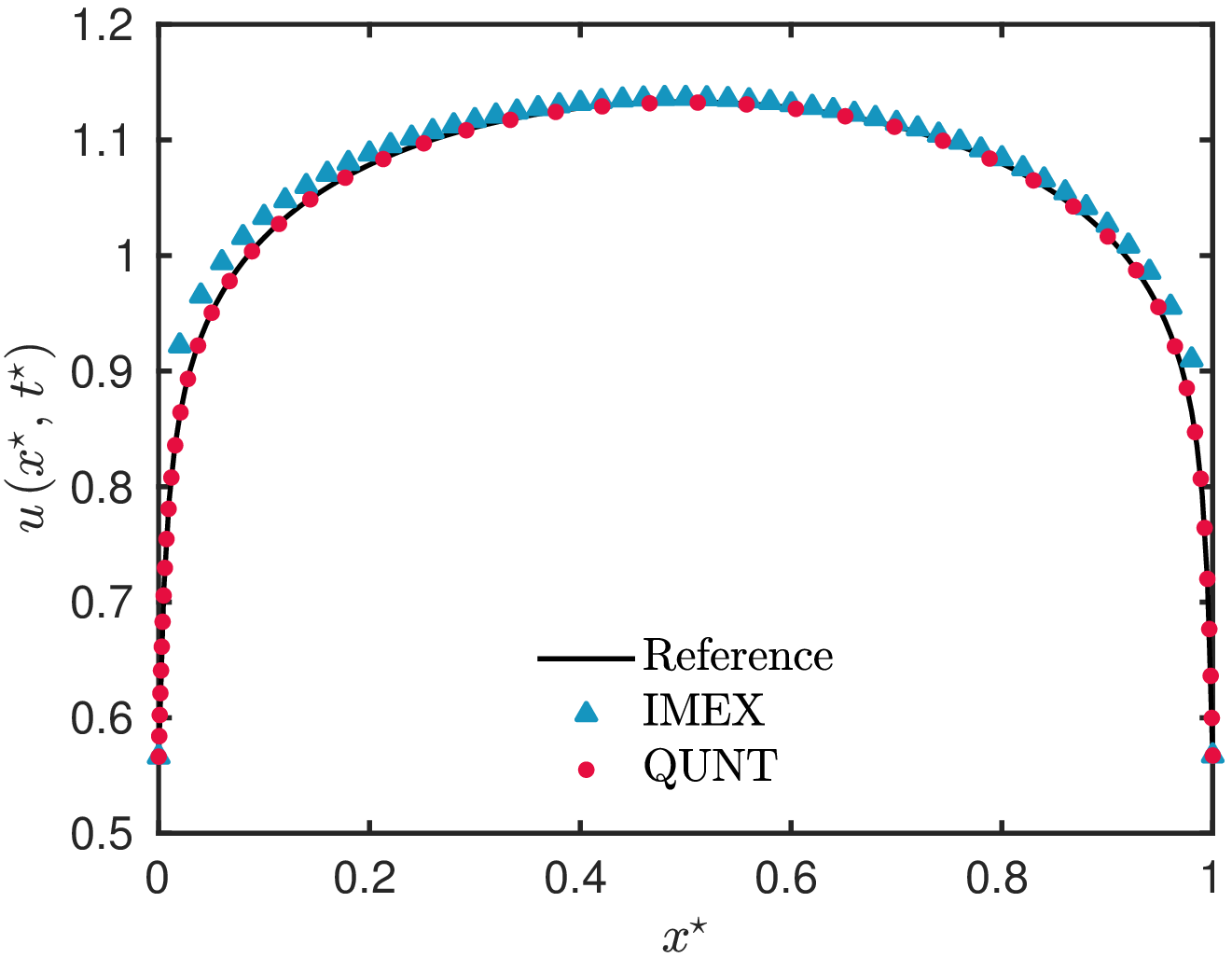}}
\subfigure[b][\label{Figure:AN1_error_prof}]{\includegraphics[width=0.48\textwidth]{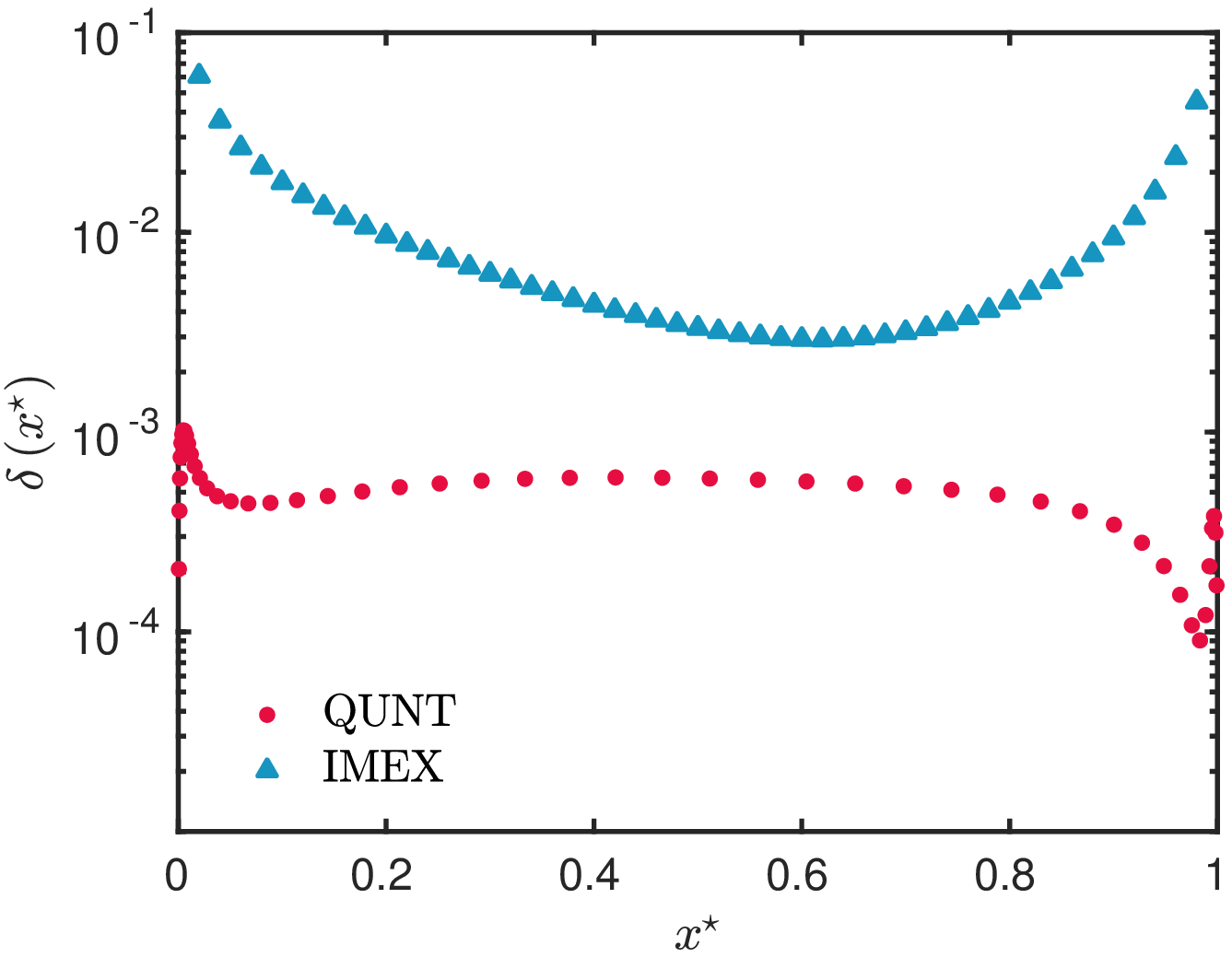}}
\caption{\small\em Temperature profiles for the  time $\ts\ =\ 16$ (a) and the error $\delta\,(\xs)$ for the same profile (b).}
\label{Figure:AN1_prof}
\end{figure}

Figure~\ref{Figure:AN1_profile} presents the temperature profile for $\ts \egalb 16$ of the computed solutions. As the material has a low thermal conductivity, the temperature in the interior of the material does not change as fast as it does at near the boundaries. Consequently, high gradients can be observed on the boundaries, making the spatial nodes concentrate in these areas. The QUNT method handles much better these high gradients than the IMEX, providing more accurate results. As the QUNT method was built based on the IMEX scheme, we can conclude that the moving grid modifications improved the numerical scheme. The absolute error $\delta$ for this instant of time is presented in Figure~\ref{Figure:AN1_error_prof}. This figure shows that the QUNT has a spatial accuracy the order of $\O\,(10^{\,-\,4})\,$, while the IMEX scheme has an accuracy the order of $\O\,(10^{\,-\,3})\,$. Thus for the same quantity of spatial nodes $(N_{\,x} \egalb 51)\,$, the QUNT can be $10$ times more accurate than the numerical solution of the IMEX scheme. If we look at the boundaries, this ratio is even higher reaching $\approx \, 100$ times.

\begin{figure}
\centering
\includegraphics[width=0.5\textwidth]{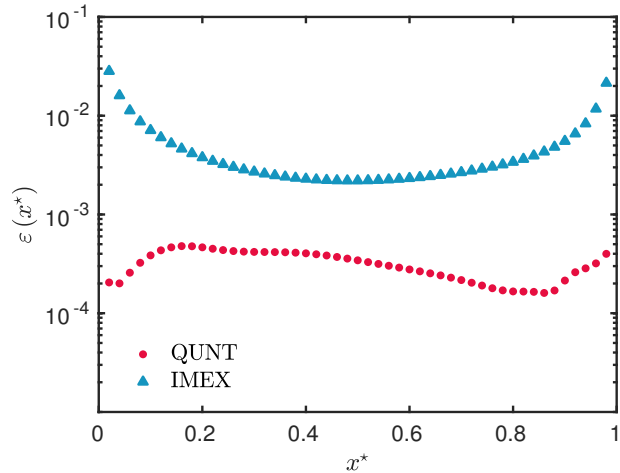}
\caption{\small\em The error $\varepsilon\,(x)$ over the whole simulation.}
\label{Figure:AN1_err_time}
\end{figure}

The results for the error over the time are presented in Figure~\ref{Figure:AN1_err_time}. The order of the error remains stable for the whole simulation, with a maximum error of $\varepsilon_{\,\infty} \,=\, 4.78\cdot 10^{\,-\,4}$ for the QUNT and $\varepsilon_{\,\infty} \,=\, 2.82\cdot 10^{\,-\,2}$ for the IMEX method, which confirms $\approx \, 100$ ratio between two approaches.

After simulating the temperature field, it is of great importance to get the values of the heat flux density at the boundaries, to compute, for instance, the transmission loads. Therefore, in what follows, we study the influence of the number of spatial nodes on the heat flux density.

\begin{figure}
\centering
\subfigure[a][\label{Figure:AN1_fluxL2}]{\includegraphics[width=0.48\textwidth]{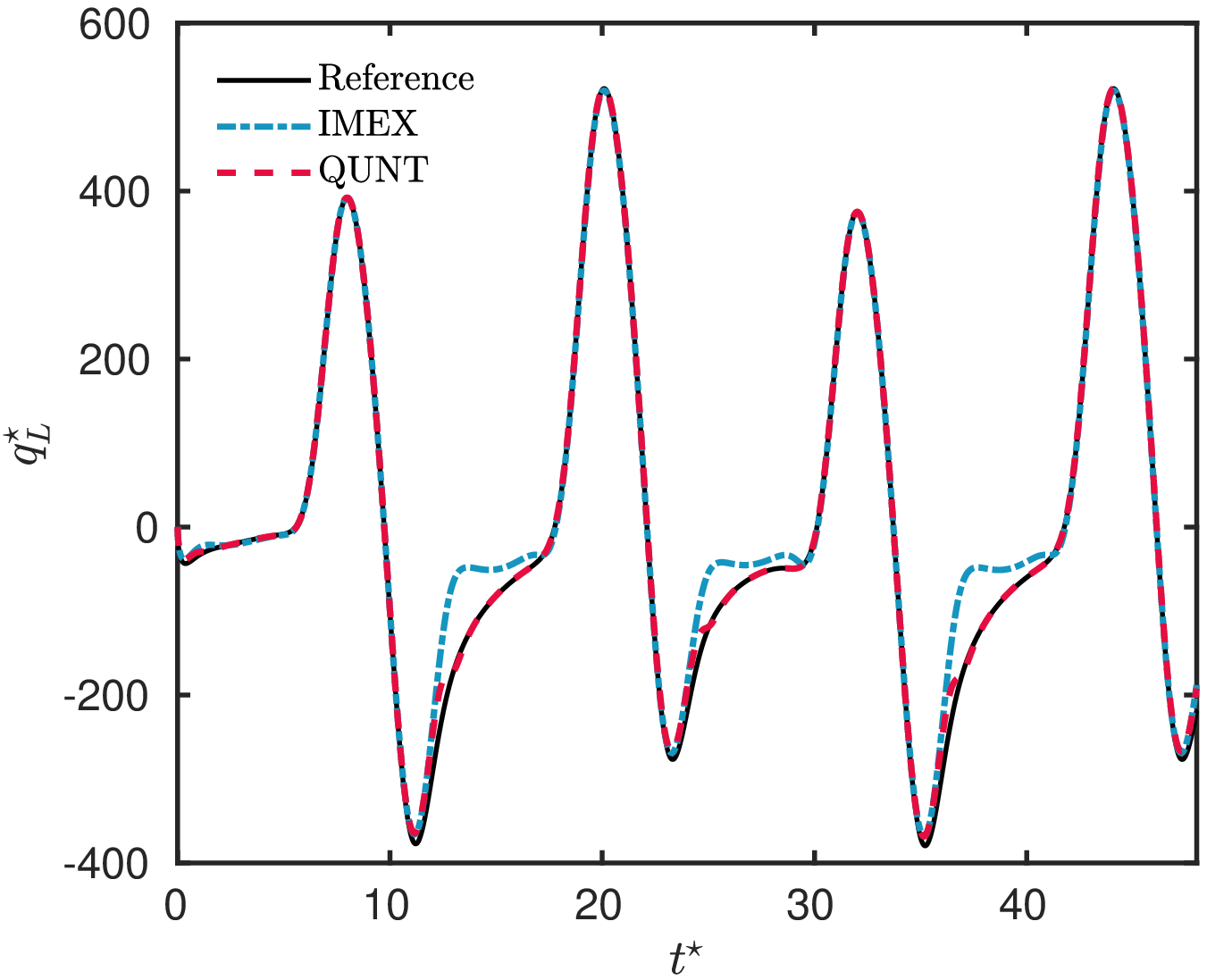}}
\subfigure[b][\label{Figure:AN1_fluxR2}]{\includegraphics[width=0.48\textwidth]{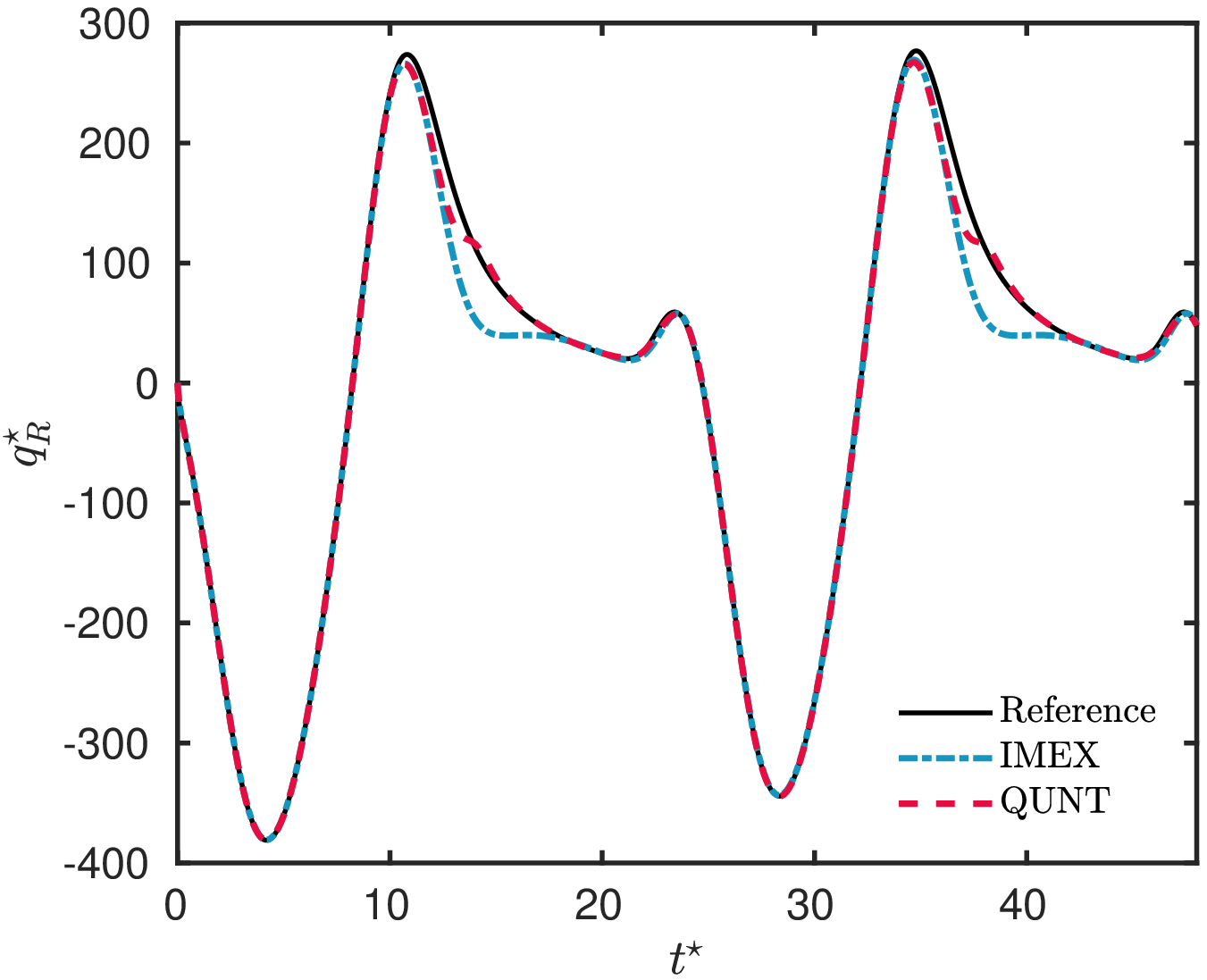}}
\caption{\small\em Flux evolution computed at the left (a) and at the right (b) boundaries.}
\label{Figure:AN1_flux}
\end{figure}

Figures~\ref{Figure:AN1_fluxL2} and \ref{Figure:AN1_fluxR2} present the heat flux evaluated at the left and right boundaries, respectively, with the derivative approximated with a finite-difference scheme second-order accurate in space. The results show that there is a difference between the flux computed for the uniform grid with the one computed for the non-uniform grid.

\begin{figure}
\centering
\subfigure[a][\label{Figure:AN1_flux_errL}]{\includegraphics[width=0.48\textwidth]{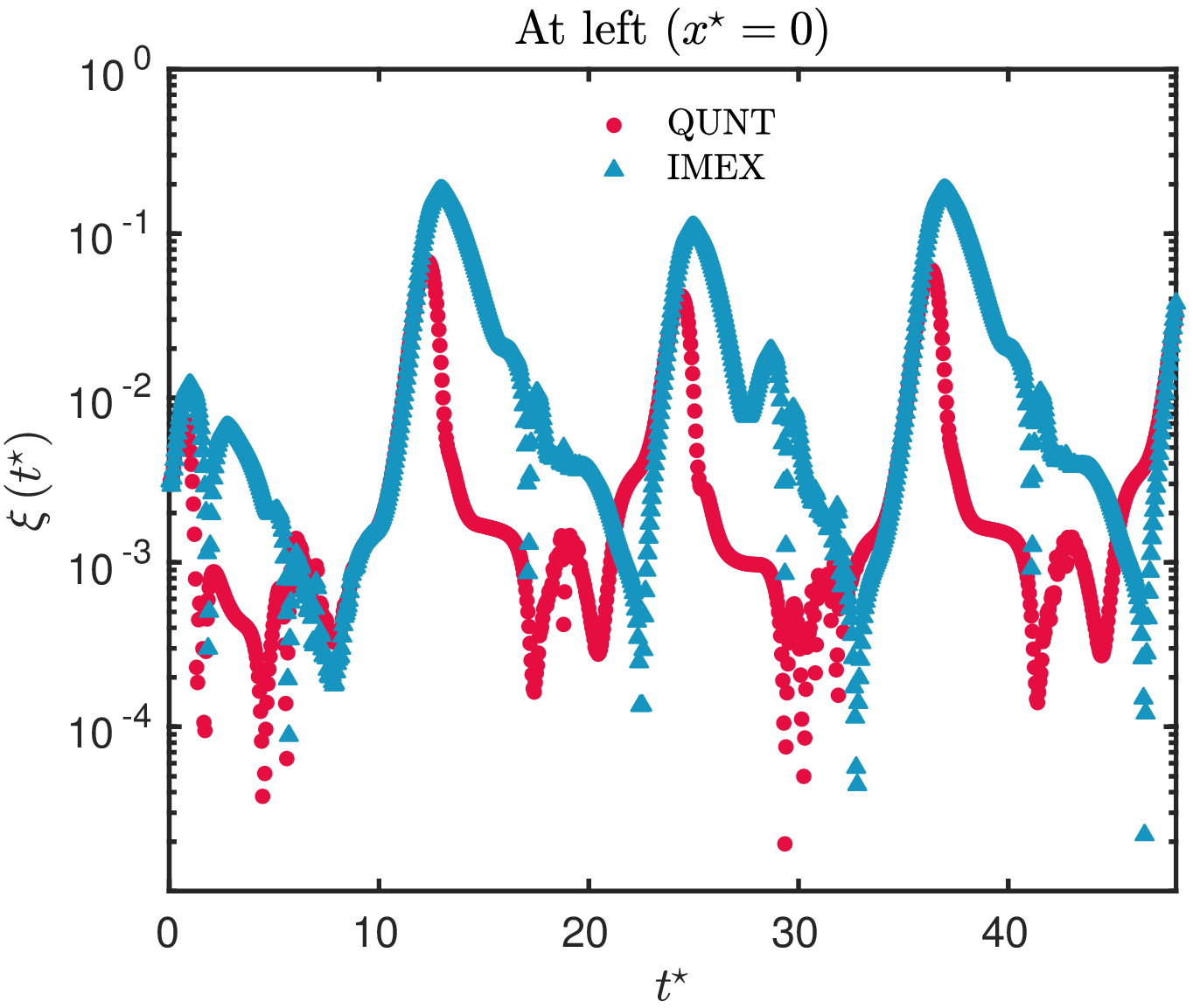}} \hspace{0.3cm}
\subfigure[b][\label{Figure:AN1_flux_errR}]{\includegraphics[width=0.48\textwidth]{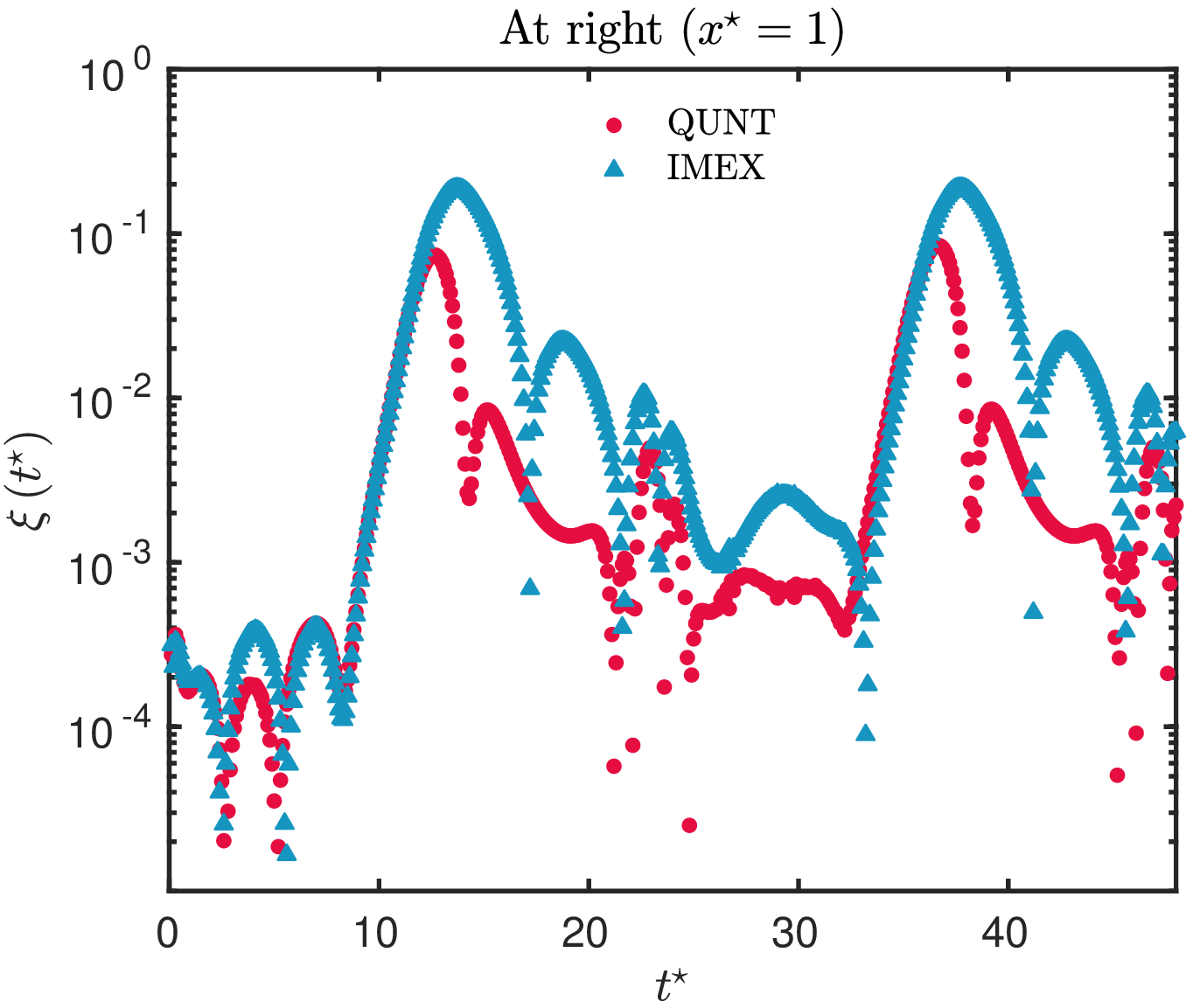}}
\caption{\small\em Flux error $\xi$ computed at the left boundary (a) and at the right boundary (b).}
\label{Figure:AN1_flux_err}
\end{figure}

Figures~\ref{Figure:AN1_flux_errL} and \ref{Figure:AN1_flux_errR} represent the error of the fluxes $\xi\,$, computed with the QUNT and IMEX solutions for the left and right boundaries, respectively. In general, the flux computed with the non-uniform grid is more accurate than the flux computed with the uniform grid. By using $51$ spatial nodes, the maximum error of the flux computed with an uniform grid is $\xi_{\,\infty}\egalb 1.9\cdot 10^{\,-\,1}$ and the flux computed with a non-uniform grid is $\xi_{\,\infty}\egalb 8.5\cdot 10^{\,-\,2}\,$.

\subsection{Convergence study}

\begin{figure}
\centering
\subfigure[a][\label{Figure:AN1_err_dt1}]{\includegraphics[width=0.48\textwidth]{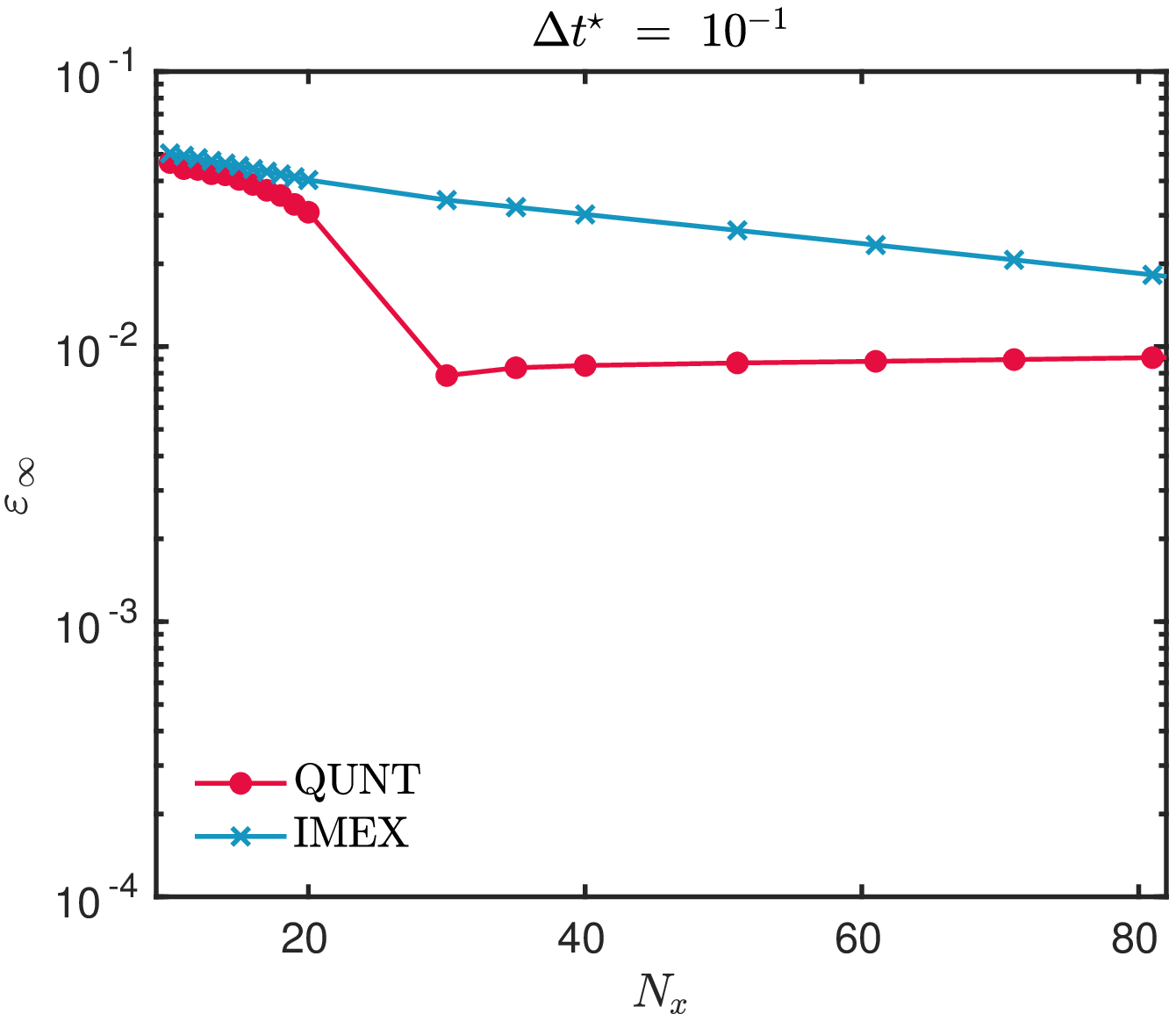}}
\subfigure[b][\label{Figure:AN1_err_dt2}]{\includegraphics[width=0.48\textwidth]{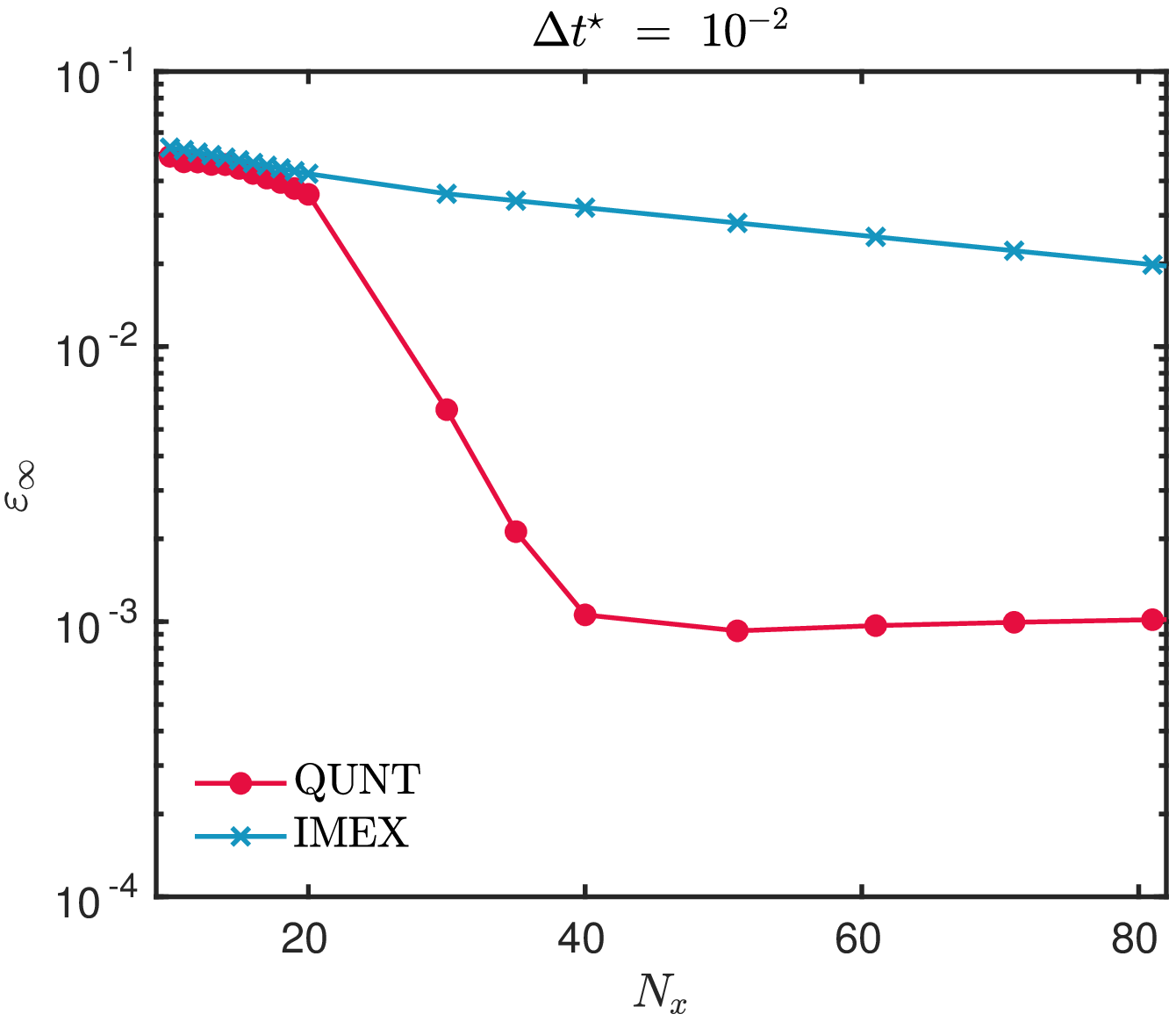}}\\
\subfigure[c][\label{Figure:AN1_err_dt3}]{\includegraphics[width=0.48\textwidth]{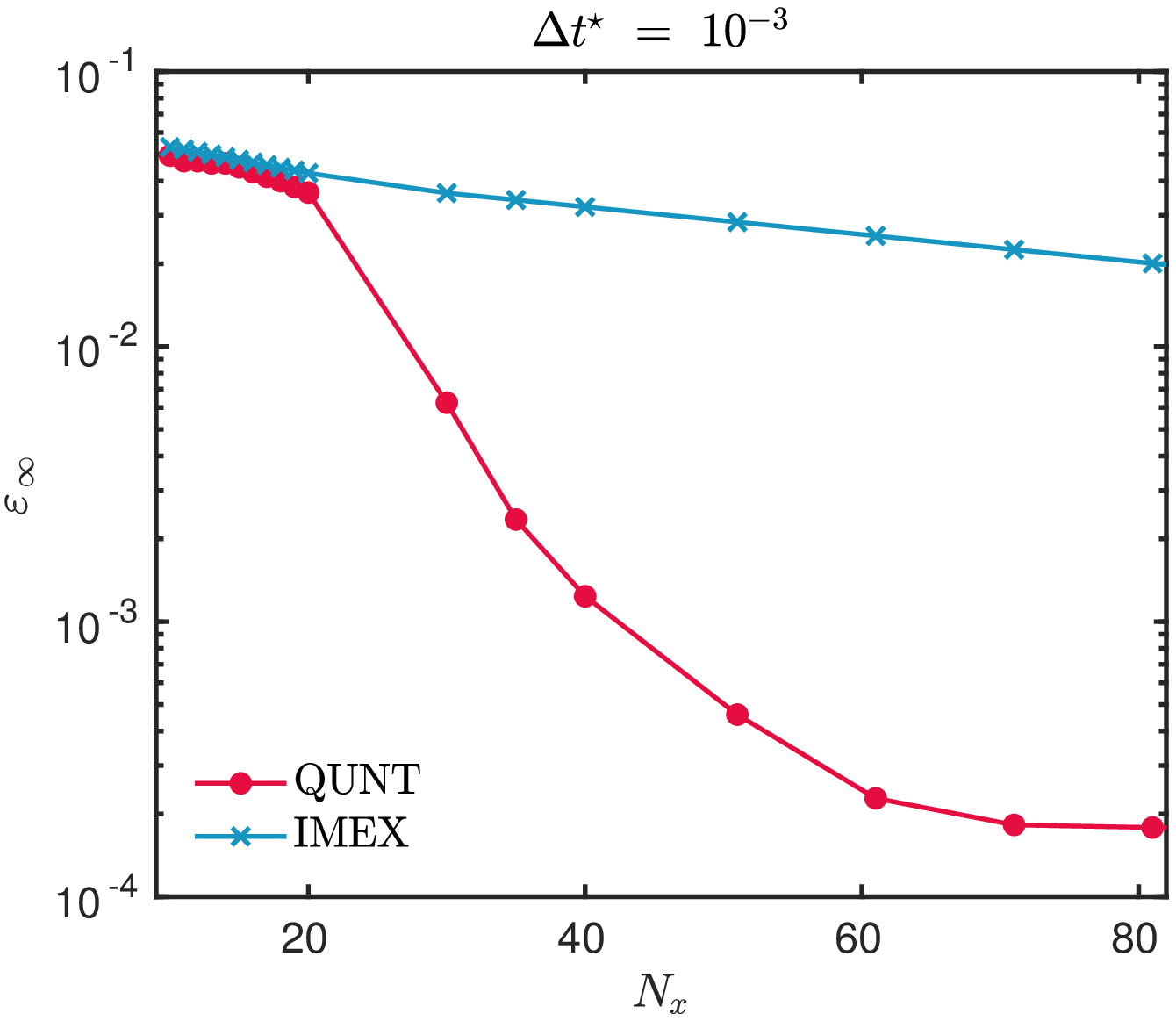}}
\subfigure[c][\label{Figure:AN1_err_dt4}]{\includegraphics[width=0.48\textwidth]{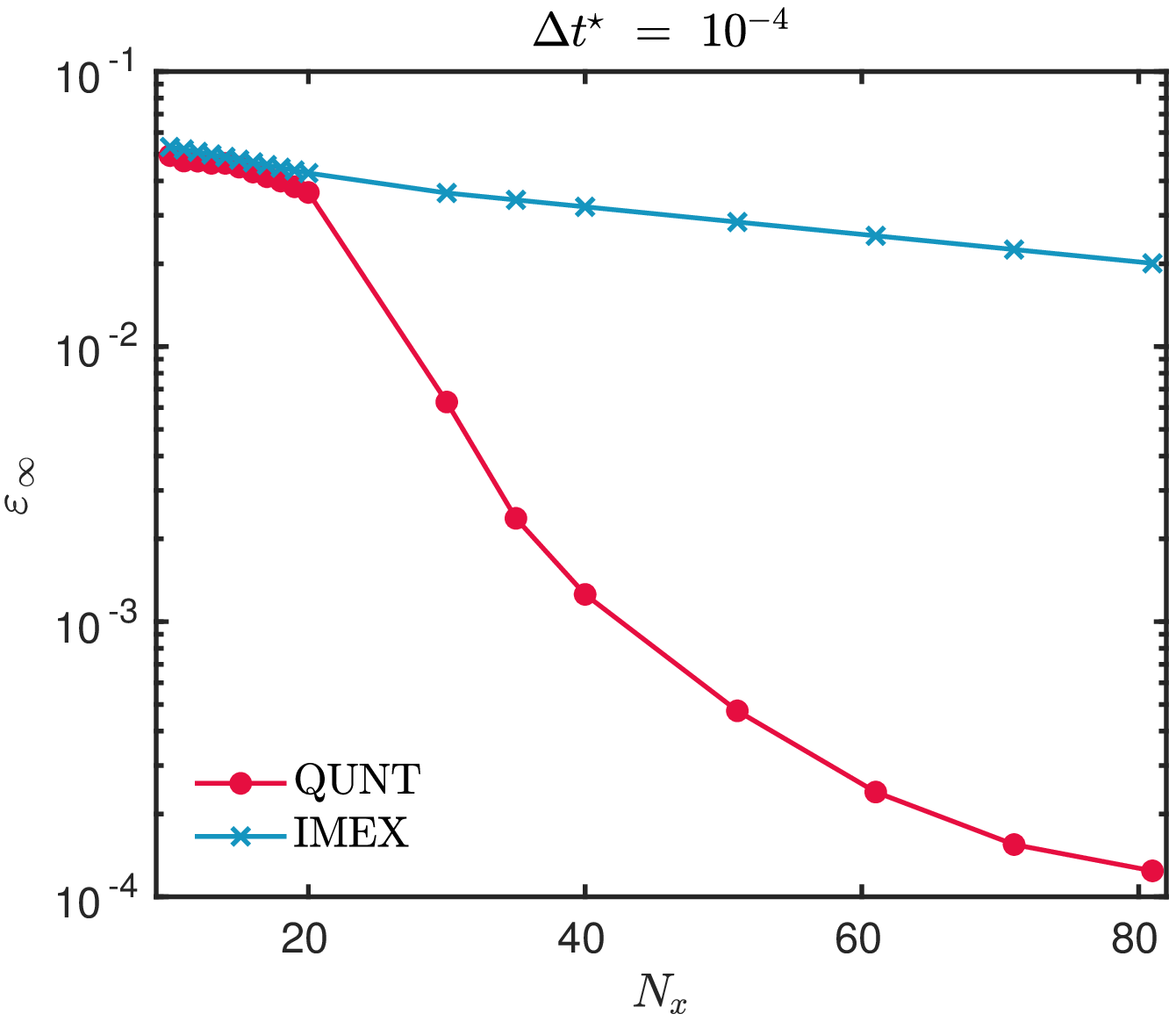}} 
\caption{\small\em Global error of the temperature field $\varepsilon_{\,\infty}$ computed in function of the number of spatial nodes, for $\dts \egalb 10^{\,-\,1}$(a), for $\dts \egalb 10^{\,-\,2}$(b), for $\dts \egalb 10^{\,-\,3}$(c), and for $\dts \egalb 10^{\,-\,4}$(d).}
\label{Figure:AN1_param}
\end{figure}

A convergence study was carried out to compare the QUNT with the IMEX scheme and to analyze the accuracy of both temperature distribution field and heat flux density. Thus, for different values of $\dts \egalb \bigl\{\,10^{\,-\,1}\,; 10^{\,-\,2}\,; 10^{\,-\,3}\,; 10^{\,-\,4}\,\bigr\} \,$, the global error $\varepsilon_{\, \infty}$ was computed as a function of the number of spatial nodes $N_{\,x} \ \in \ [\,10\,,80\,]$. Figures~\ref{Figure:AN1_err_dt1}, \ref{Figure:AN1_err_dt2}, \ref{Figure:AN1_err_dt3}, and \ref{Figure:AN1_err_dt4} display the results.

It can be seen in Figure~\ref{Figure:AN1_param} that for $N_x\ < \ 20$ both methods provide solutions of the field with approximately the same order of accuracy, which is insufficient to get good approximations of the solution. For the IMEX method, less than $100$ nodes are not enough to give a solution with an accuracy better than $\O\,(10^{\,-\,2})\,$. Even if the $\dts$ is decreased, the global error is still of the same order of accuracy due to insufficient spatial discretization. On the other hand, for the QUNT method, as the number of spatial nodes increase, the global error decreases faster than the IMEX method. For the QUNT adaptive grid, with only $40$ nodes, an accuracy of $\O\,(10^{\,-\,3})$ can be achieved for the three values of $\dts\,$.

\begin{figure}
\centering
\subfigure[a][\label{Figure:AN1_parm_flux_dt1}]{\includegraphics[width=0.48\textwidth]{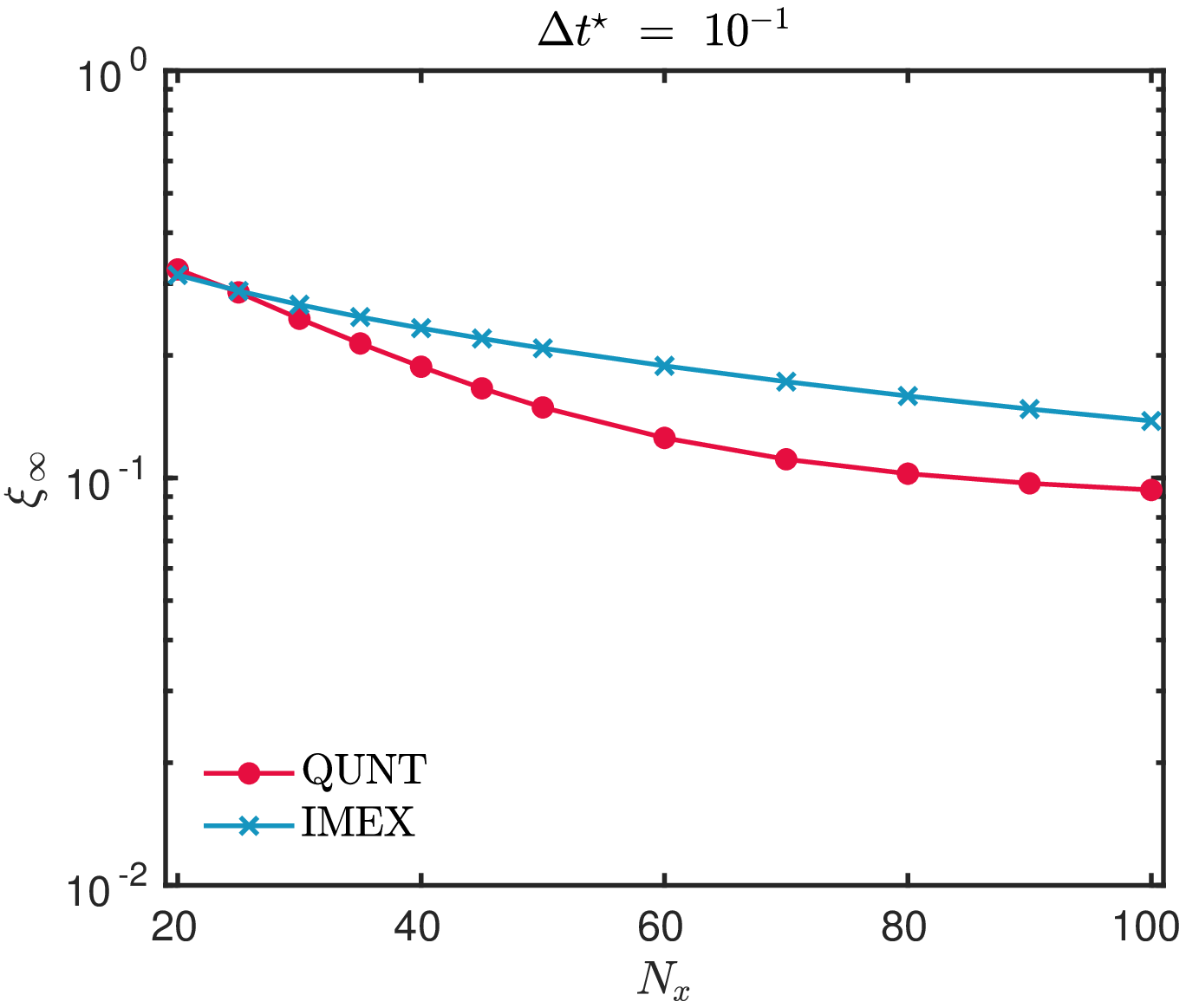}} \hspace{0.3cm}
\subfigure[b][\label{Figure:AN1_parm_flux_dt2}]{\includegraphics[width=0.48\textwidth]{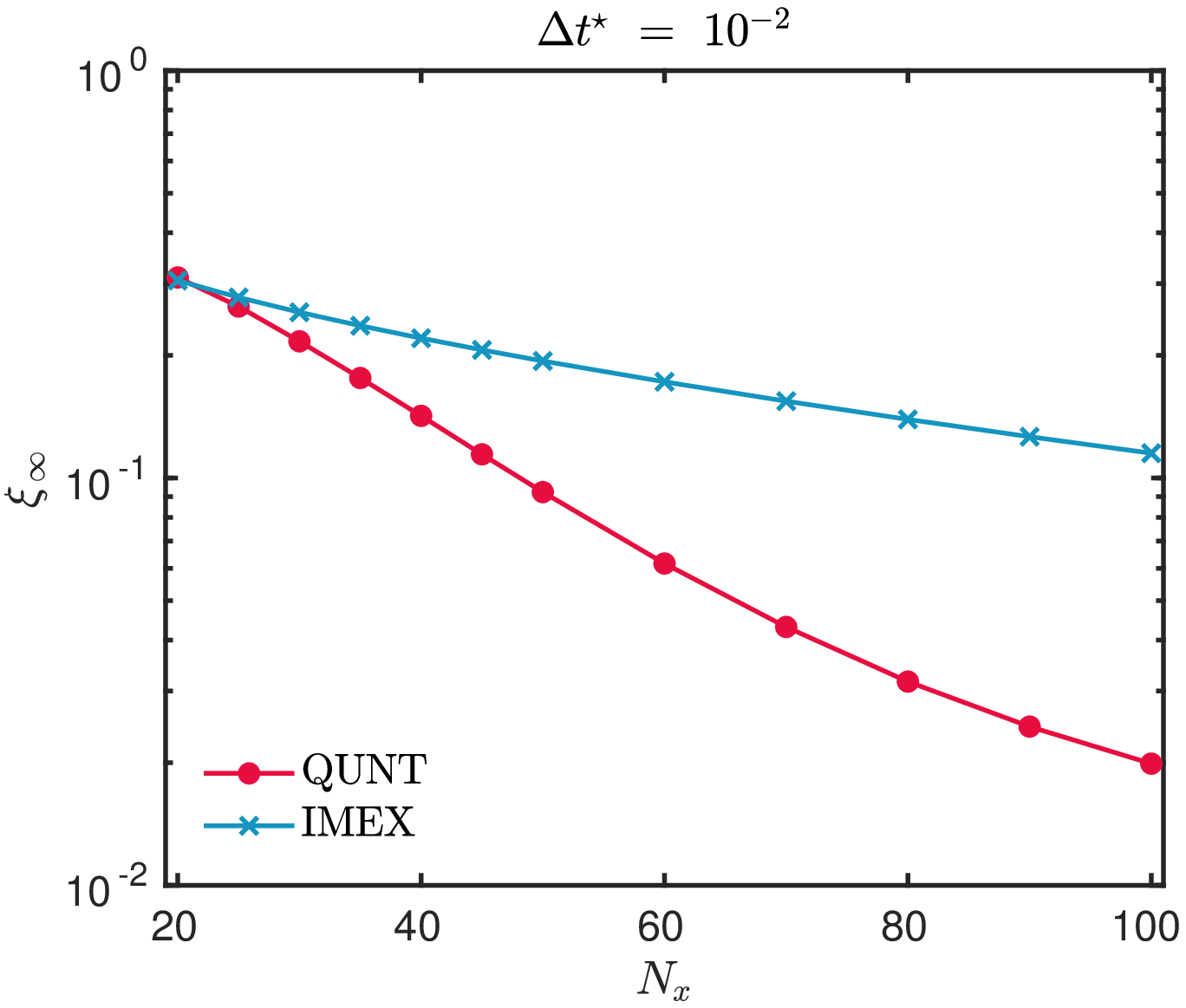}}\\
\subfigure[c][\label{Figure:AN1_parm_flux_dt3}]{\includegraphics[width=0.48\textwidth]{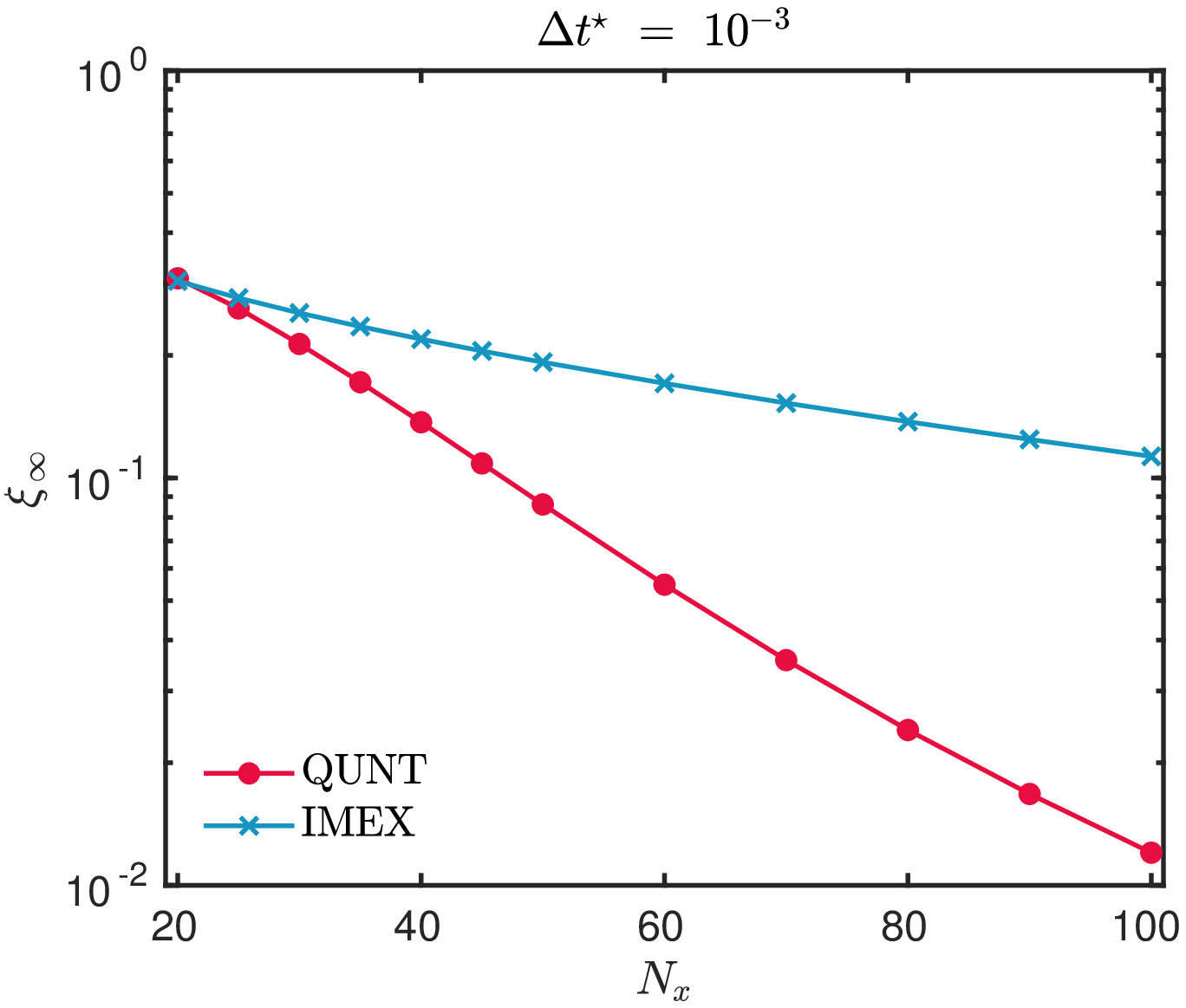}} \hspace{0.3cm}
\subfigure[d][\label{Figure:AN1_parm_flux_dt4}]{\includegraphics[width=0.48\textwidth]{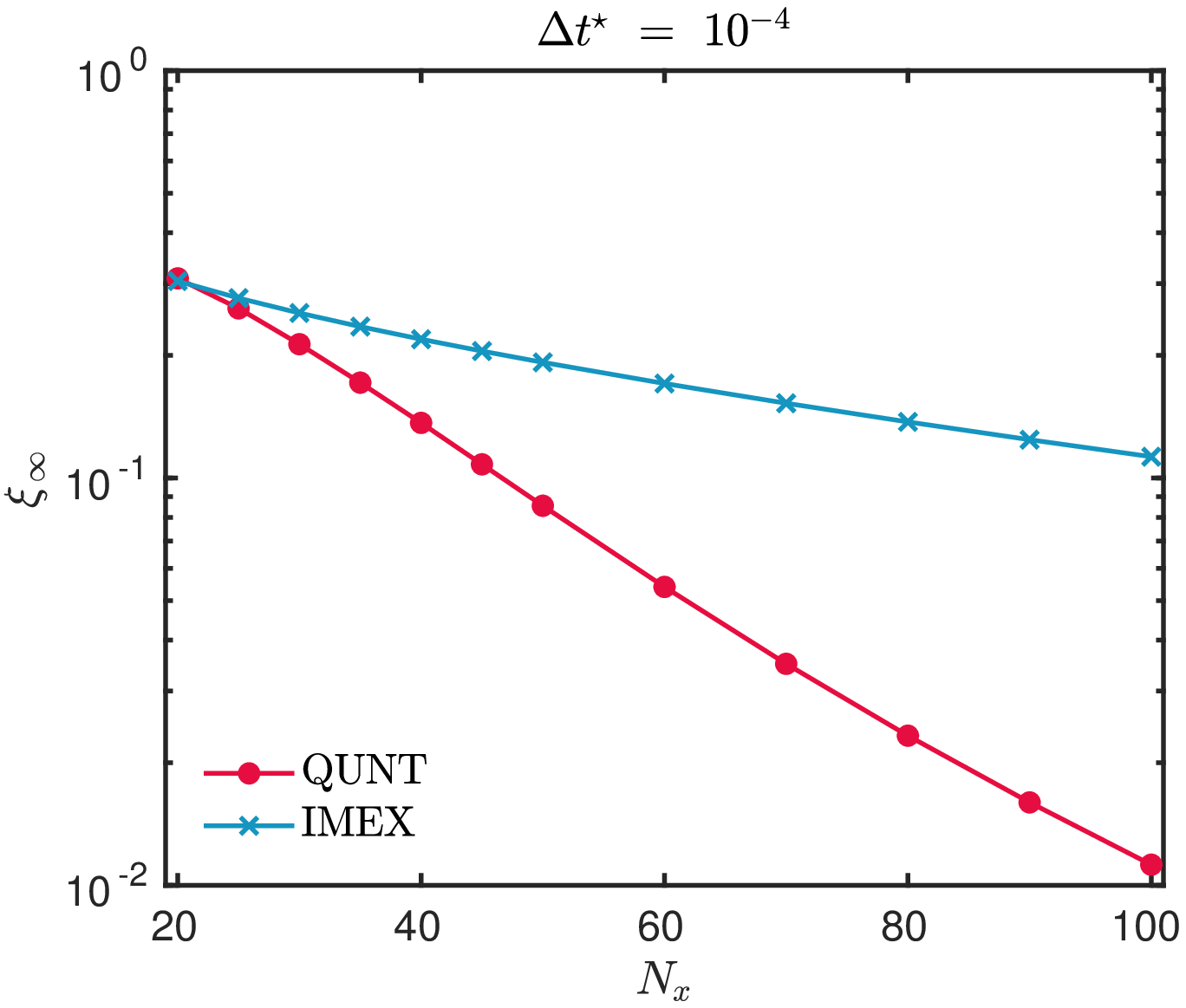}} 
\caption{\small\em Global error of the heat flux density $\xi_{\,\infty}$ computed as a function of the number of spatial nodes, for $\dts \egalb 10^{\,-\,1}$(a), for $\dts \egalb 10^{\,-\,2}$(b), for $\dts \egalb 10^{\,-\,3}$(c) and for $\dts \egalb 10^{\,-\,4}$(d).}
\label{Figure:AN1_parm_flux}
\end{figure}

Figures~\ref{Figure:AN1_parm_flux_dt1}, \ref{Figure:AN1_parm_flux_dt2}, \ref{Figure:AN1_parm_flux_dt3} and \ref{Figure:AN1_parm_flux_dt4} display the results for the global error of the flux density $\xi_{\, \infty}\,$, as a function of the number of spatial nodes $N_{\,x} \ \in \ [\,20\,,100\,]$ also for different values of $\dts \egalb \{\,10^{\,-1}\,; 10^{\,-2}\,; 10^{\,-3}\,; 10^{\,-4}\,\}$. Results show that the computation of the flux density is less accurate than the computation of the temperature field. This happens since the error increases when the derivative has to be computed. The QUNT method also provides solutions for the flux more accurate than the IMEX method. As the analysis of the error for the field, the QUNT approach proved to be more sensible regarding variations of $\dt$, as the solution becomes smoother and the moving grid adapts better when $\dt$ decreases. Although, the same accuracy is observed for $\dt \egalb 10^{\,-\,3}$  and $\dt \egalb 10^{\,-\,4}$ because the order of the error is also related with the spatial discretization. To obtain higher accuracy, it is necessary to increase the  derivative approximation order. In this case, the error of the flux is relatively high due to the nonlinearity of the case study. For a less nonlinear case, the error of the flux density can be considerably reduced.

Therefore, it has been clearly noticed the QUNT approach computes an accurate solution of the problem. Moreover, the moving grid method converges to an accurate solution with less spatial nodes than the method with an uniform grid. In a long-term simulation, it does provide significant gains in regard to the CPU usage. The computer used to assess the computational cost has a processor Intel\textsuperscript\textregistered Core\textsuperscript{TM}i$5$ @ $2.80$GHz$\times 4$ with $15.6$GB of RAM.

\begin{table}
\center
\caption{\small\em Computational time required for the numerical schemes to perform the nonlinear case for $\varepsilon_{\,\infty} \cong \O (10^{\,-\,4})\,$.}
\bigskip
\small
\setlength{\extrarowheight}{.3em}
\begin{tabular}[l]{@{}cccc}
\hline
\textit{Simulation time} & \textit{Crank-Nicolson} & \textit{QUNT} & \textit{IMEX} \\
\hline
$2$ days    &  $1.43\ \mathsf{s}$   & $1.12\ \mathsf{s}$  & $1.08\ \mathsf{s}$ \\
$1$ month &  $21.36\ \mathsf{s}$  & $16.2\ \mathsf{s}$  & $15.2\ \mathsf{s}$ \\
$1$ year  &  $259.32\ \mathsf{s}$ & $205.1\ \mathsf{s}$ & $185.9\ \mathsf{s}$ \\
\hline
\textit{Spatial nodes} &  $1001$ & $51$ & $501$ \\
\hline
\textit{CPU time ($\% $)} &  $100\%$ & $75\% $  & $72\% $ \\
\hline
\end{tabular}
\label{tab:CPU_time}
\end{table}

To evaluate the gains in terms of computational cost, Table~\ref{tab:CPU_time} provides the CPU time to compute the solution using the \textsc{Crank--Nicolson}, the QUNT and the IMEX methods, which have been evaluated using the \texttt{Matlab\texttrademark} platform. All methods computed a solution with the same order of accuracy $\varepsilon_{\,\infty} \cong \O\,(10^{\,-\,4})\,$. For this, it was used a time step of $\dts \egalb 10^{\,-\,2}\,$, and $51$ spatial nodes for the QUNT and $1\,001$ and $501$ spatial nodes for the \textsc{Crank--Nicolson} and IMEX methods, respectively. The computational effort to perform the simulation increases linearly with the simulation time. The QUNT and IMEX methods can compute the solution $25\%$ faster than the \textsc{Crank--Nicolson} approach. It is preferable to focus on the ratio of computer run-time rather than on absolute values, since it is system-dependent. The \textsc{Crank--Nicolson} scheme was implemented with a fixed point approach to treat the non-linearity and it was used here only for comparison purposes. One should notice that the codes are not optimized, and the differences between the CPU time of QUNT and IMEX methods are low, even with the QUNT method performing two extra matrix inversions.


\section{Optimum insulation in Brazil}
\label{sec:application}

In the previous section, the QUNT method was presented and validated. Now, the method is used to compute the solution of a transient one-dimensional case study of heat transfer through building walls for a period of one year. The results are obtained with less computational efforts and enable to estimate the optimum thermal insulation for large cities in \textsc{Brazil}.


\bigskip
\paragraph{Climatic zones.}

\textsc{Brazil} has a total surface area of $8.5$ million $\mathsf{km^2}$, the fifth largest country in the world, occupying almost half of the entire \textsc{South America} continent. In such a large country a variety of climates exist. However, most of the country can be defined as being tropical and sub-tropical and energy consumption for space conditioning is dominated by cooling. Although, in the \textsc{Southern} region of \textsc{Brazil}, energy is also used for heating during the winter months (\textsc{June -- August}).

For this study, four \textsc{Brazilian} cities --- \textsc{Curitiba}, \textsc{Rio de Janeiro}, \textsc{S\~ao Paulo} and \textsc{Salvador} --- have been selected for determining their optimal insulation thickness, chosen due to the diversity of the climate and due to their large population \cite{Morishita2016}. However, the method used in this study can be applied to any city desired. Based on ASHRAE Standard $90.1-2013$ \cite{ASHRAE2013}, these cities are located in three different climatic zones: \textit{Very Hot-Humid} (zone $1$), \textit{Hot-Humid} (zone $2$) and \textit{Warm-Humid} (zone $3$). Table~\ref{Table:climate} presents some characteristics of the chosen cities.

\begin{table}
\centering
\caption{\small\em Climate and geographic characteristic of the \textsc{Brazilian} cities.}
\bigskip
\setlength{\extrarowheight}{.3em}
\begin{tabular}{ccccccc}
\hline
City   & \begin{tabular}[c]{@{}c@{}}$T_{\min}$\\ $[\oC]$\end{tabular} & \begin{tabular}[c]{@{}c@{}}$T_{\text{mean}}$\\ $[\oC]$\end{tabular} & \begin{tabular}[c]{@{}c@{}}$T_{\max}$\\ $[\oC]$\end{tabular} & \begin{tabular}[c]{@{}c@{}}Altitude\\ $[\mathsf{m}]$ \end{tabular} & Location & Clim. zone\\
\hline
\textsc{Curitiba}       & $-2 $   & $16.3$ & $30.9$ & $935$ &  $25.5^{\circ}$ S $49.3^{\circ}$ W  & $3$  \\
\textsc{Rio de Janeiro} & $13$    & $23.5$ & $38.2$ & $2$   &  $22.9^{\circ}$ S $43.2^{\circ}$ W  & $1$  \\
\textsc{S\~ao Paulo}      & $7.5$   & $18.8$ & $32.8$ & $760$ &  $23.5^{\circ}$ S $46.6^{\circ}$ W  & $2$ \\
\textsc{Salvador}       & $14.2$  & $25.3$ & $33.5$ & $8$   &  $12.9^{\circ}$ S $38.5^{\circ}$ W  & $1$ \\
\hline
\end{tabular} 
\label{Table:climate}
\end{table}

\bigskip
\paragraph{Simulation parameters.}

\begin{figure}
\centering
\includegraphics[width=0.9\textwidth]{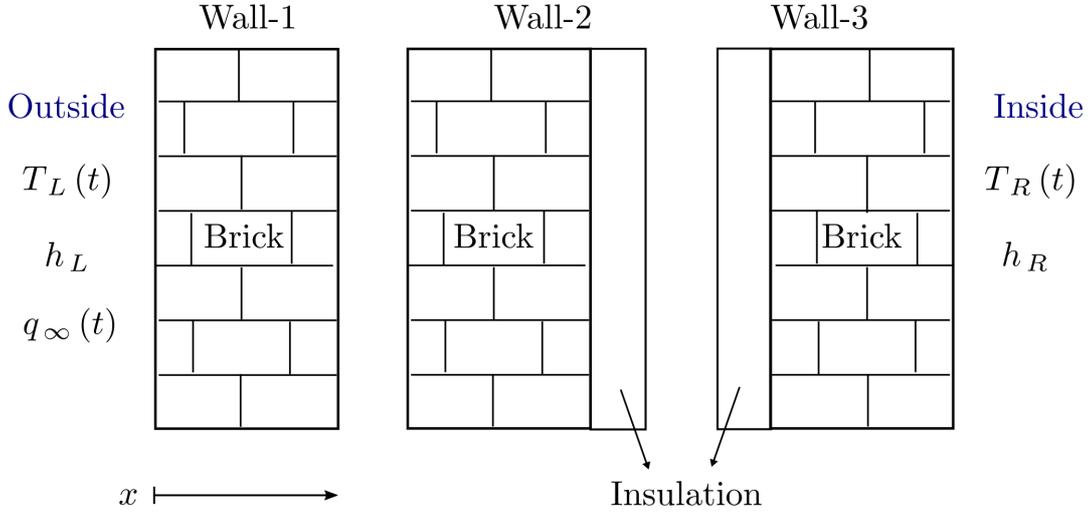}
\caption{\small\em Schematic representation of the wall's composition.}
\label{Figure:wall_model}
\end{figure}

Bricks are commonly used in typical wall structures in \textsc{Brazilian} constructions \cite{Morishita2016, Civil2003}. As illustrated in Figure~\ref{Figure:wall_model}, we consider $3$ types of walls: Wall$-1$ is composed only of brick; Wall$-2$ includes also a layer of insulation on the internal side; Wall$-3$ has a layer of insulation on the external side. The brick is $15\ \mathsf{cm}$ thick, while the thickness of insulation material (extruded polystyrene) varies. The thermal properties of the materials are gathered from the IEA Annex$-24$ \cite{KumarKumaran1996} and are reported in Table~\ref{Table:properties}.

\begin{table}
\centering
\caption{\small\em Thermophysical properties of the materials \cite{KumarKumaran1996}.}
\bigskip
\setlength{\extrarowheight}{.3em}
\begin{tabular}{c c c c}
\hline
\textit{Material} & $\rho_{\,0}\, \mathsf{[m^3/kg]}$ & $c_{\,p}\, \mathsf{[J/kg/K]}$ & $k\, \mathsf{[W/m/K]}$ \\
\hline
Brick & $1800$ & $840$ & $0.69$ \\
Concrete & $2200$ & $840$ & $2$ \\
Extrude Polystyrene & $25$ & $1470$ & $0.0275$ \\
\hline
\end{tabular}
\label{Table:properties}
\end{table}

\begin{figure}
\centering
\subfigure[a][\label{Figure:BC_temp_cwb}]{\includegraphics[width=0.48\textwidth]{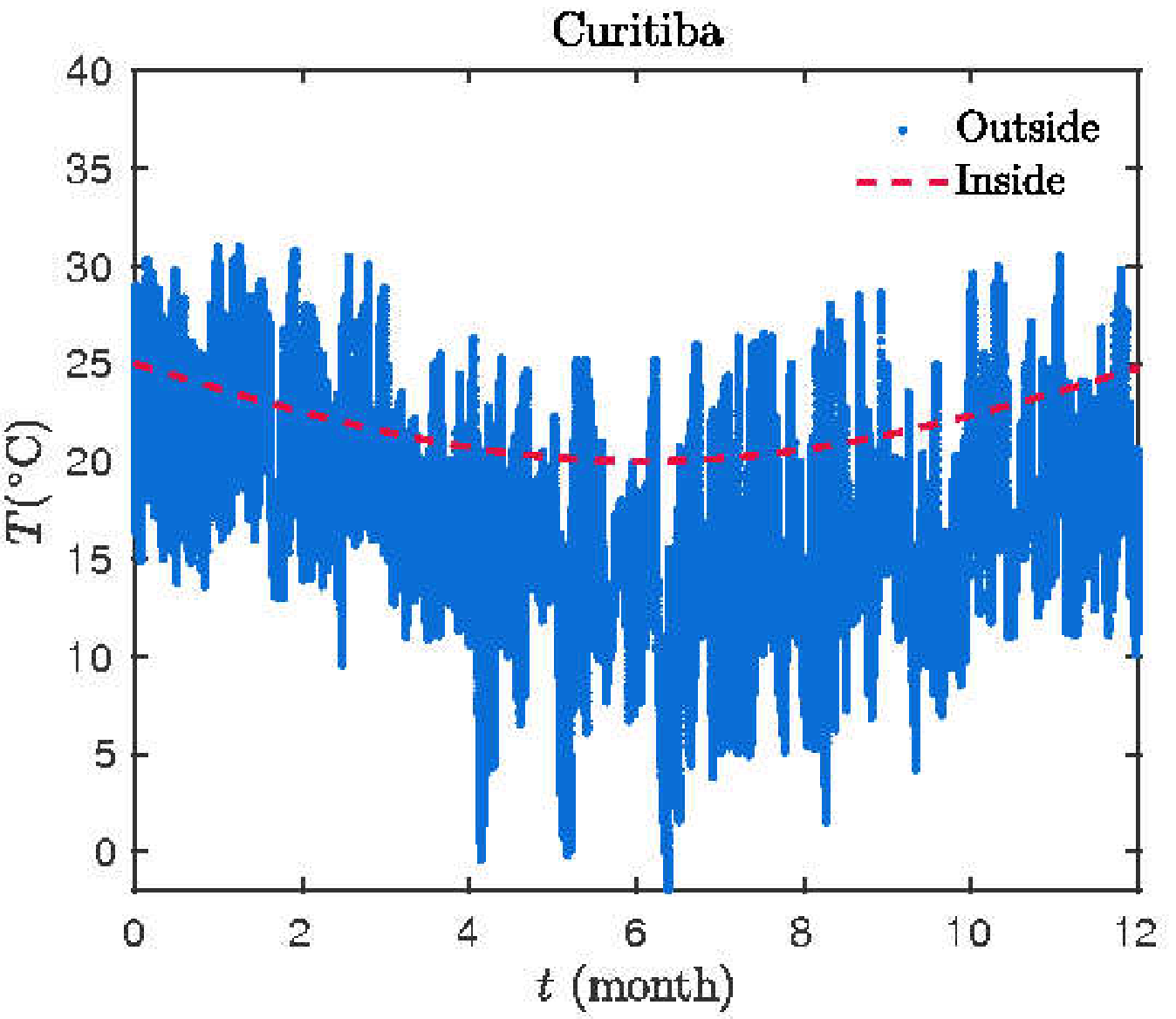}}
\subfigure[b][\label{Figure:BC_temp_rio}]{\includegraphics[width=0.48\textwidth]{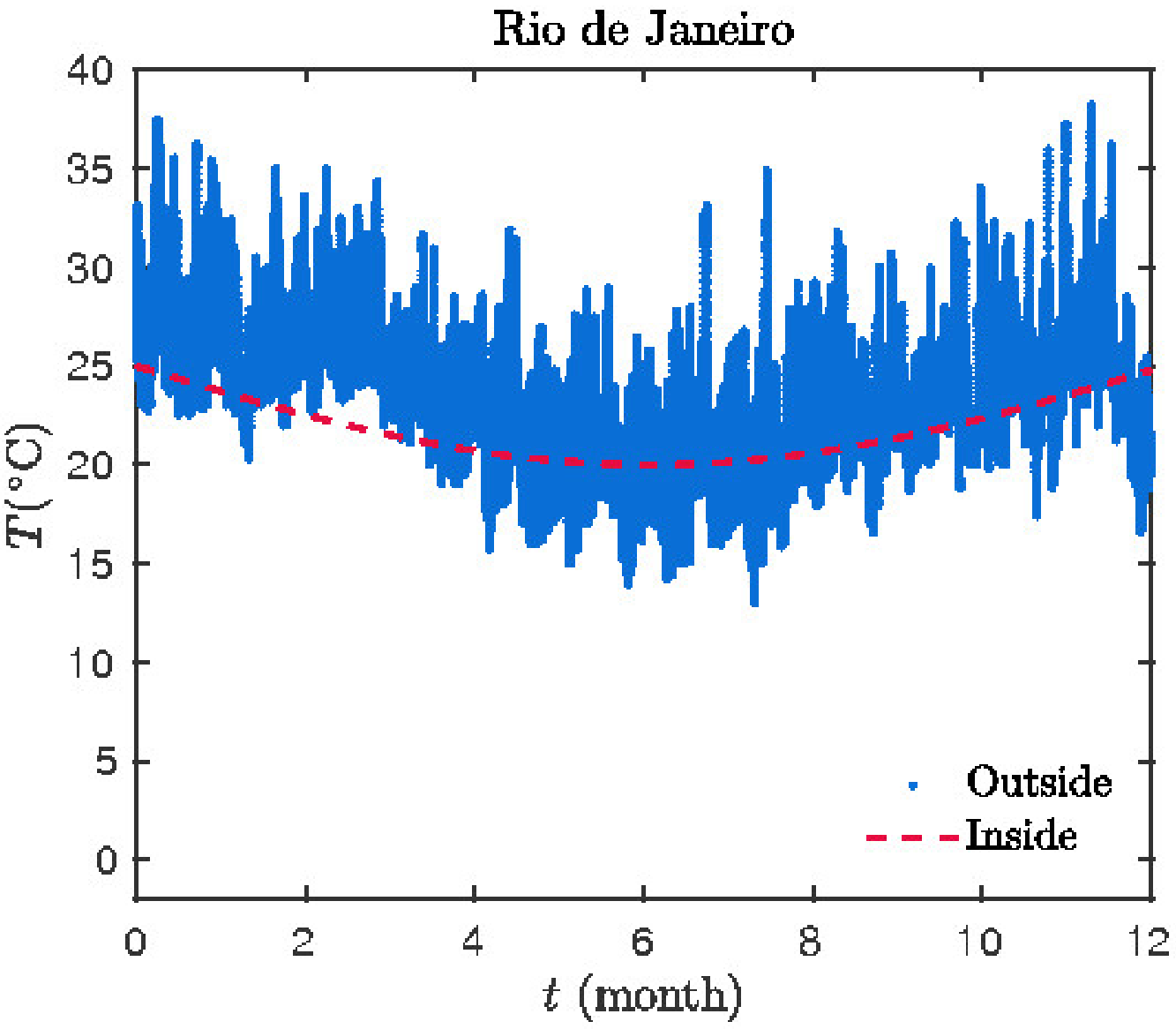}}
\subfigure[c][\label{Figure:BC_temp_sp}]{\includegraphics[width=0.48\textwidth]{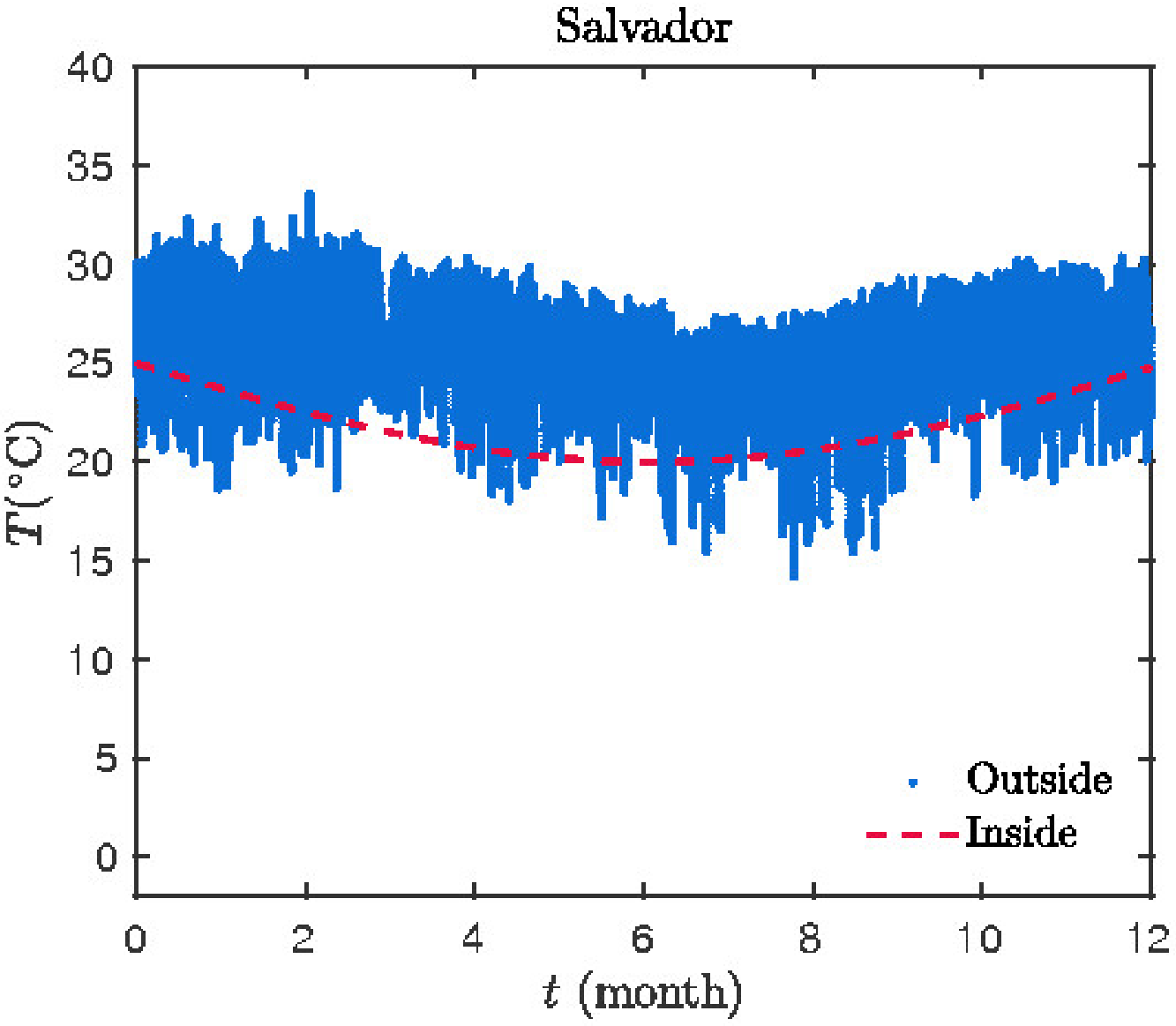}}
\subfigure[d][\label{Figure:BC_temp_sal}]{\includegraphics[width=0.48\textwidth]{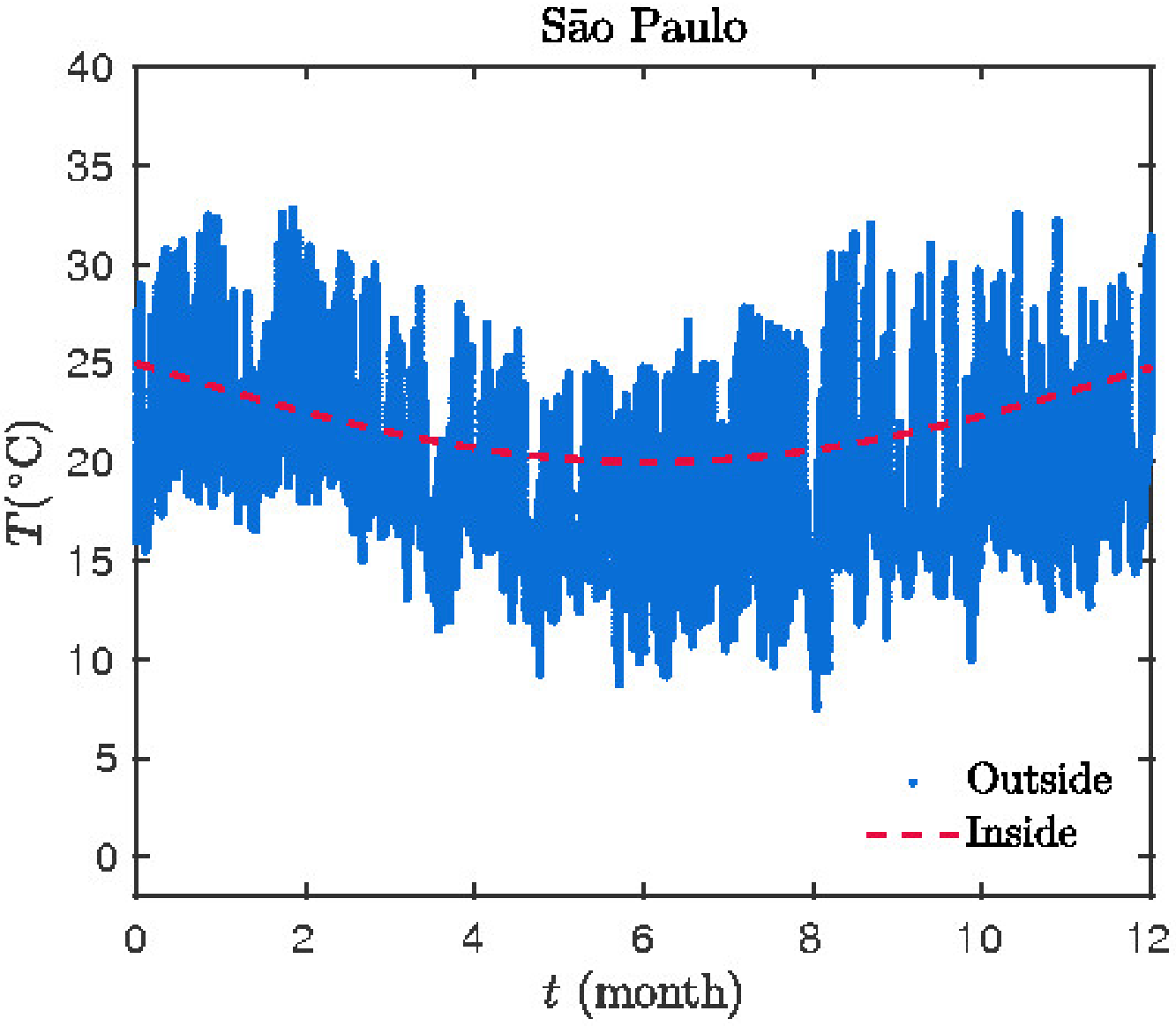}} 
\caption{\small\em Inside and outside air temperatures for \textsc{Curitiba} (a), \textsc{Rio de Janeiro} (b), \textsc{Salvador} (c) and \textsc{S\~ao Paulo} (d).}
\label{Figure:BC_temp}
\end{figure}

\begin{figure}
\centering
\includegraphics[width=0.75\textwidth]{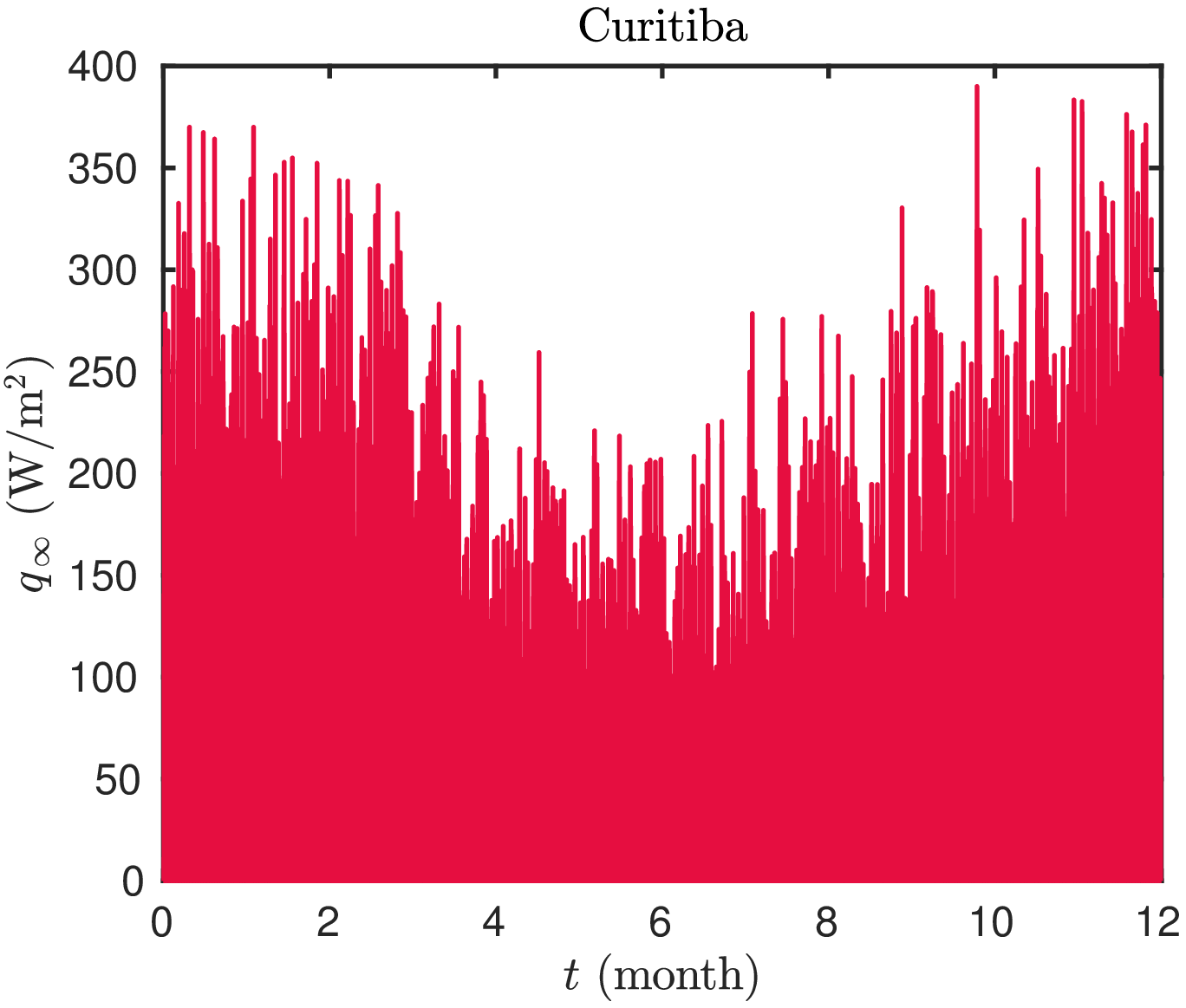}
\caption{\small\em The heat flux incidence on the \textsc{South} wall fa\c{c}ade.}
\label{Figure:BC_flux_cwb}
\end{figure}

As the initial condition, a steady-state temperature distribution within the wall is assumed:
\begin{align*}
  \Ti\,(x) \egal \dfrac{T_R(0) \moins T_L(0)}{l}\, x \plus T_L(0) \quad [\mathsf{^{\,\circ}C}]\,.
\end{align*}
Climatic conditions of temperature and solar radiation are considered on the outside part of the wall for a period of one year. The outside temperature and total radiation are given in Figures~\ref{Figure:BC_temp} and \ref{Figure:BC_flux_cwb}. The outside heat flux is computed using radiation data from \texttt{Domus} \cite{Mendes2008} for a \textsc{South}-oriented wall. The solar absorptivity of the outside surface is taken as $\alpha \egalb 0.5\,$. The indoor air temperature $\TR(t)$  has a sinusoidal variation as illustrated in Figure~\ref{Figure:BC_temp}. During the summer period, the temperature is set to $25\mathsf{^{\circ}C}$ and during the winter to $20\mathsf{^{\circ}C}$ \cite{Melo2014}. The convective heat transfer coefficients at the outer and the inner surfaces are set to constant values $\hL \egalb 25 \ \mathsf{W/(m^{\,2}\cdot K)}$ and $\hR \egalb 10 \ \mathsf{W/(m^{\,2}\cdot K)}$, respectively. The long-wave radiative transfer on the external surface is only considered for the roof simulation.

According to Figure~\ref{Figure:BC_temp}, the inside temperature in all cities varies in the same way. What can be observed, is that  \textsc{Curitiba} is the coldest city, which means it needs more heating than cooling. In \textsc{Rio de Janeiro} and \textsc{Salvador}, the external temperatures are higher than the inside temperature, which means that the cooling system predominates. In addition, \textsc{Salvador} has a higher mean external temperature than \textsc{Rio de Janeiro}. \textsc{S\~ao Paulo} has the mildest weather among the four cities, but the external temperatures are mostly lower than the internal ones, which makes the heating need prevails.

In next sections, the thermal behavior of buildings located in these cities is studied for different wall configurations, various thicknesses of insulation material and different wall orientations, with respect to the heat flux density $q$ and transmission loads $E$ on the inside surface of the building wall. Three cases are distinguished. First, the benefits of insulating buildings are demonstrated. The second case analyzes if the insulation should be placed on the inside or outside part of the building. Last, the effect of the wall orientation is studied.

\subsection{First Case: insulated vs. non-insulated buildings}
\label{Sec:application_case1}

For the first simulation, two \textsc{South}-oriented walls in the city of \textsc{Curitiba} are considered. One has no insulation (Wall$-1$) and the other one has $10\ \mathsf{cm}$ insulation (Wall$-2$). Climatic conditions of the temperature and solar radiation are given in Figures~\ref{Figure:BC_temp_cwb} and \ref{Figure:BC_flux_cwb}.

With the numerical values provided previously, Problem~\eqref{eq:heat_equation} is solved using the QUNT method. The monitoring function is defined by Equation~\eqref{eq:monitor_function}, which has the following values for its parameters: 
$\alpha_1 \egalb 0.8\,$, $\beta_{\,1} \egalb 2\,$, $\alpha_{\,2} \egalb 0.2\,$, $\beta_{\,2} \egalb 3\,$. The grid diffusion  and the smoothing parameters are, respectively, $\beta \egalb 50$ and $\sigma \egalb 5\,$. The solution is obtained for a number of $N_{\,x} \egalb 41$ spatial nodes with a time discretization of $\dts \egalb 0.1\,$, which is equivalent to $\dt \egalb 6 \,\mathsf{min}$.

\begin{figure}
\centering
\subfigure[a][\label{Figure:flux_cwb_wall1and2}]{\includegraphics[width=0.48\textwidth]{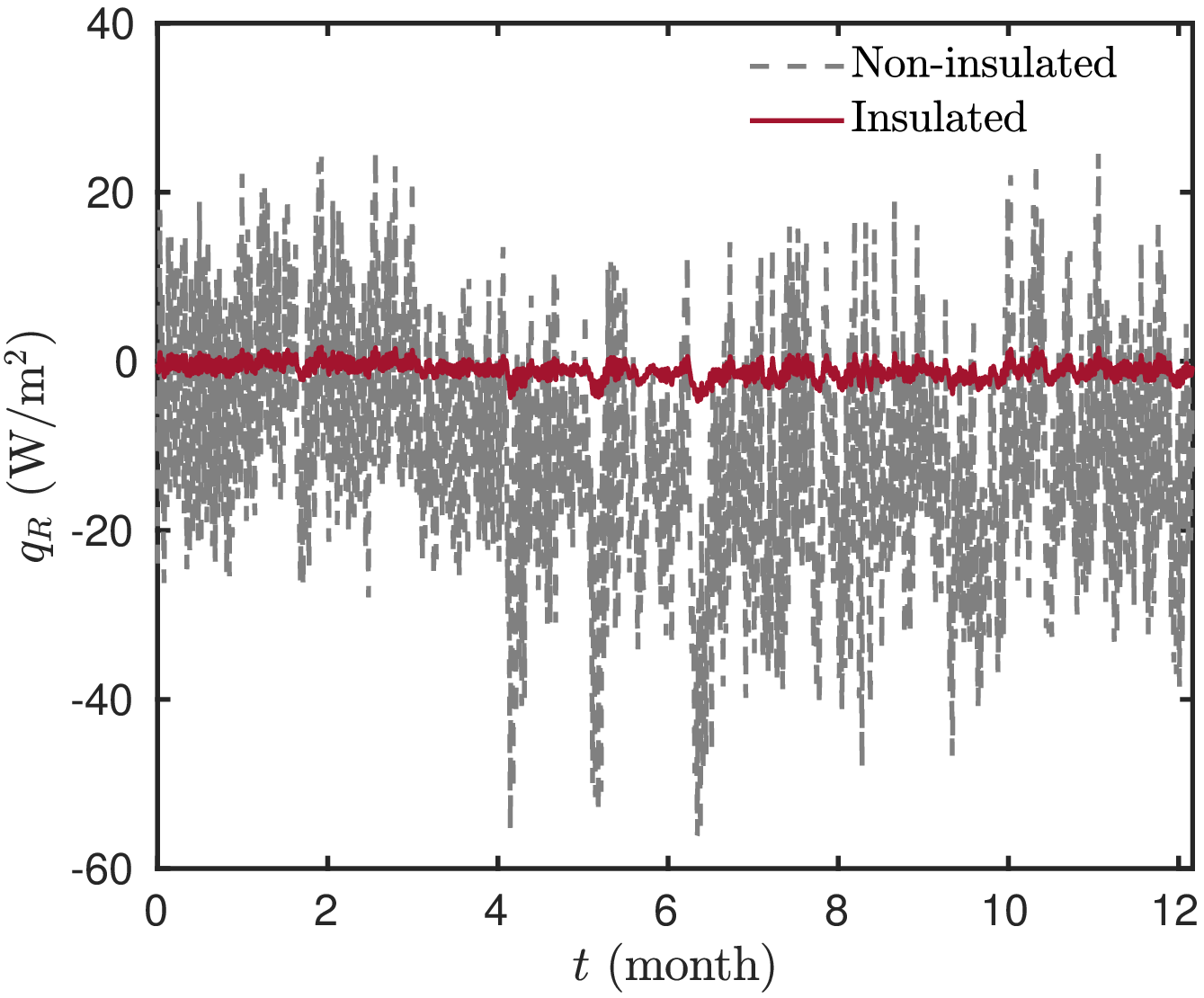}}
\subfigure[b][\label{Figure:loads_month_wall1and2}]{\includegraphics[width=0.48\textwidth]{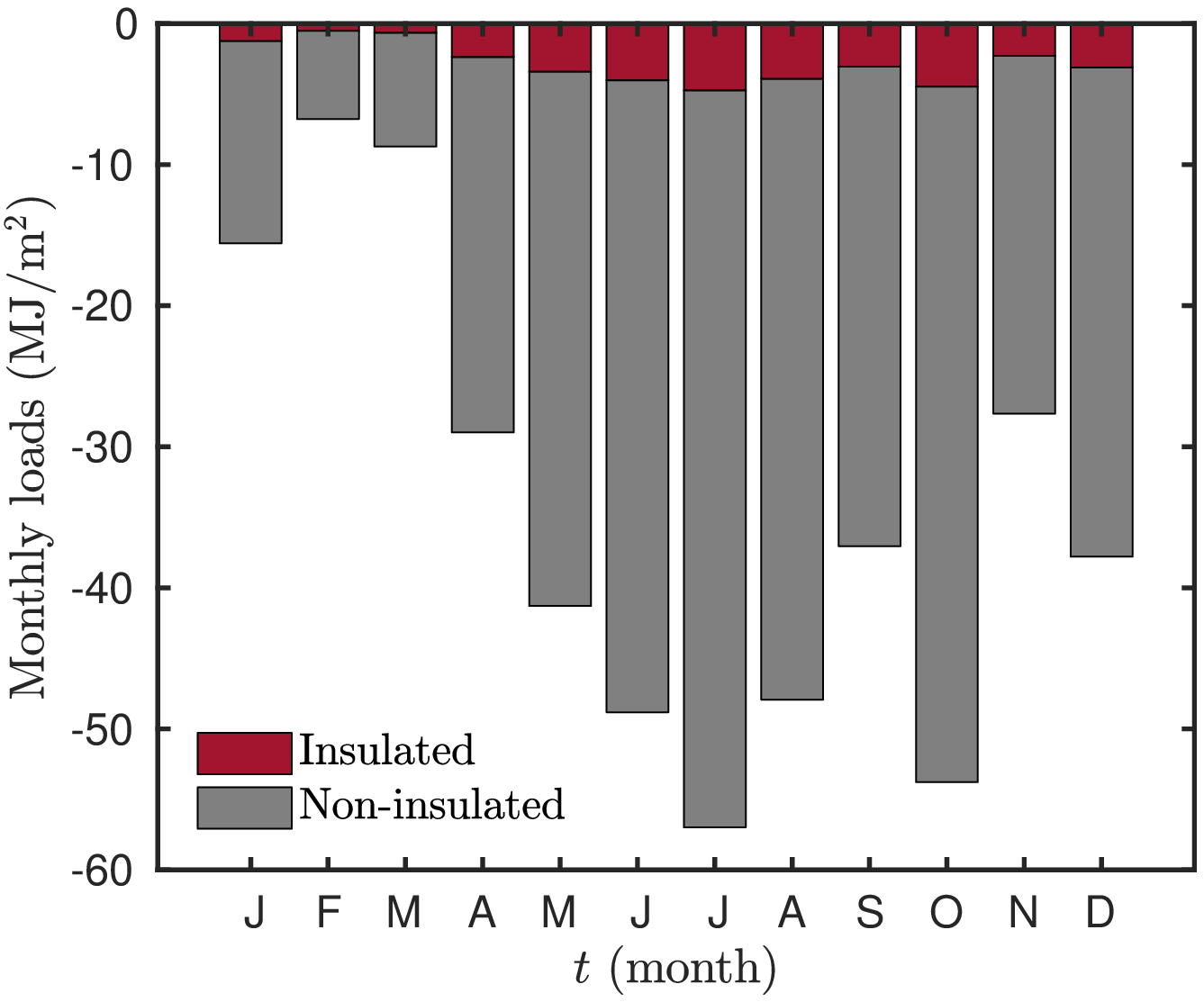}}
\caption{\small\em Heat flux density on the inside wall surface (a) and evolution of the monthly conduction load, for Wall$-1$ and Wall$-2$ models.}
\end{figure}

The temperature is computed for all points in the single- and multi-layered wall configurations. Then, the heat flux on the inside surface of the building wall is estimated and presented in Figure~\ref{Figure:flux_cwb_wall1and2}. For the case with no insulation, the heat flux has an important variation, alternating between positive and negative values. With $10\ \mathsf{cm}$ insulation, it strongly modifies the thermal behavior of the wall and the heat flux is considerably reduced in this case.

Yearly transmission loads are calculated based on Equation~\eqref{eq:load}. Hourly variation of inside surface heat flux density is integrated over a $24 \ \mathsf{h}$ period to obtain a daily total load. Then, daily total loads are integrated to get the monthly heating/cooling transmission loads presented in Figure~\ref{Figure:loads_month_wall1and2}, for Wall$-1$ and Wall$-2\,$. The heating/cooling loads are ten times higher for the wall without insulation indicating that much more energy is spent in order to maintain the temperature inside of the building.

With these results, the first conclusion is that insulation may play an important role in the construction sector in \textsc{Brazil}. In what follows, two configurations are tested, one with the insulation inside of the building wall (Wall$-2$) and another one with the insulation outside (Wall$-3$).

\subsection{Second Case: inside insulation vs. outside insulation}
\label{Sec:application_case2}

For this second case, it is also considered a \textsc{South}-oriented wall in the city of \textsc{Curitiba}, with $10\ \mathsf{cm}$ insulation. Two wall configurations are verified: one with the insulation inside the building (Wall$-2$) and another one with the insulation outside (Wall$-3$). The same climatic conditions and numerical values of the previous case (Section~\ref{Sec:application_case1}) are taken.

\begin{figure}
\centering
\subfigure[a][\label{Figure:flux_cwb_wall2and3}]{\includegraphics[width=0.48\textwidth]{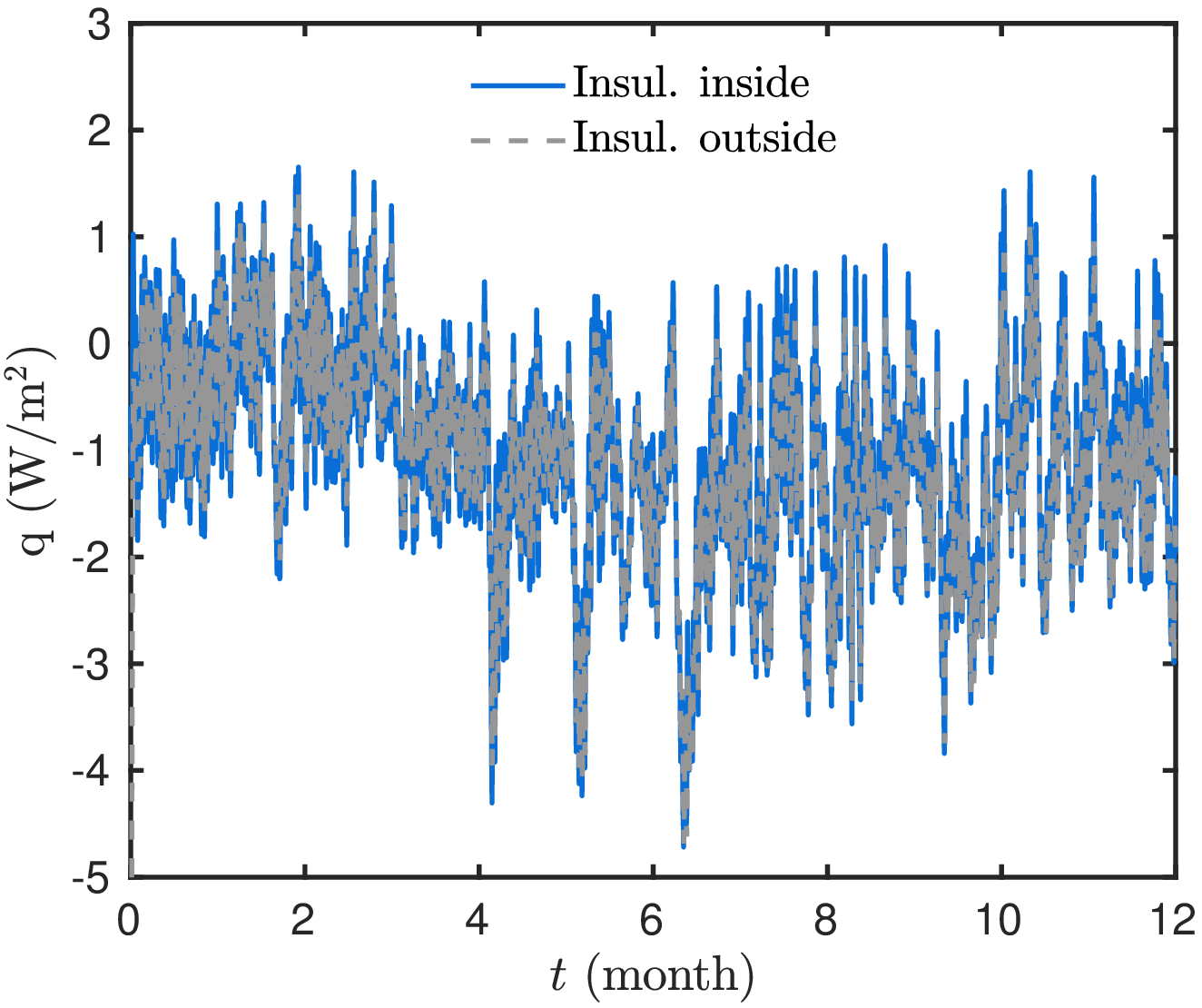}}
\subfigure[b][\label{Figure:loads_month_wall2and3}]{\includegraphics[width=0.48\textwidth]{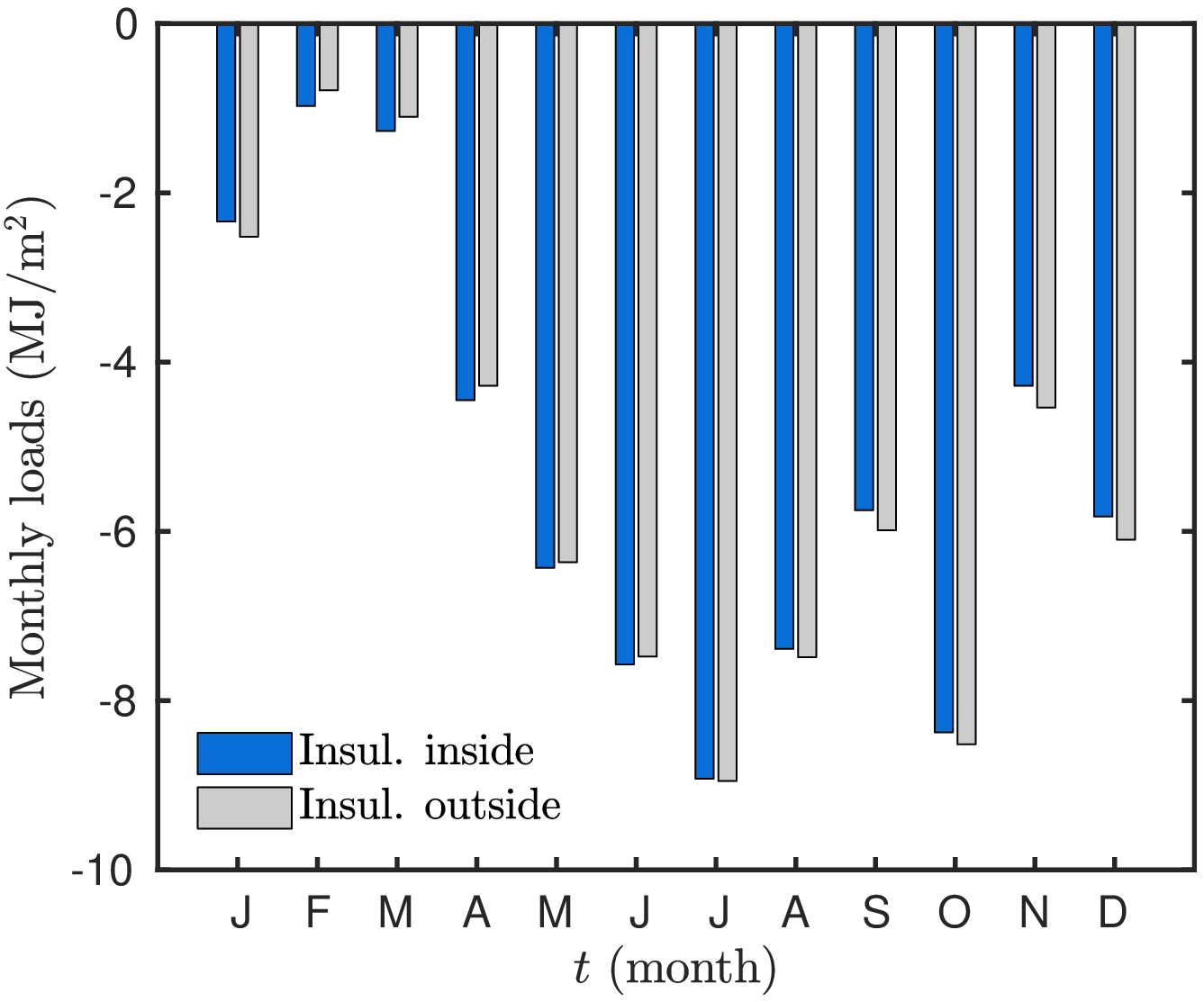}}
\caption{\small\em Heat flux density on the inside wall surface (a) and monthly conduction loads (b) for Wall$-2$ and Wall$-3$.}
\label{Figure:cwb_wall2and3}
\end{figure}

Figure~\ref{Figure:flux_cwb_wall2and3} shows the effect of the insulation position on the heat flux density during one year of simulation. The wall with external insulation has the lowest heat flux values but with a small discrepancy, with the mean difference between the heat flux of Wall-$2$ and Wall-$3$ being of the order of $0.1\, \mathsf{W/m^2}\,$, for a $10\ \mathsf{cm}$ insulation. This difference can also be noted in the monthly heating/cooling transmission loads as shown in Figure~\ref{Figure:loads_month_wall2and3}. Depending on the month, the value of the energy load can be higher for the insulation outside of the building wall. However, the annual heating/cooling transmission loads are very close, with $33.83\, \mathsf{MJ/m^2}$ for Wall$-2$ and $34.81\, \mathsf{MJ/m^2}$ for Wall$-3\,$. In \cite{Axaopoulos2014} they also have similar results for the annual loads regarding the position of the insulation (inside/outside), even with different assumptions and climatic conditions.

\subsubsection{Parametric study}
\label{Sec:application_case2_parametric}

To improve this analysis, a parametric study is performed, in which the thickness of the insulation material varies on the interval $l_{\,i} \egalb [0.01,\,0.3]\, \mathsf{m}\,$, for both wall configurations (Wall$-2$ and Wall$-3$).

\begin{figure}
\centering
\includegraphics[width=0.75\textwidth]{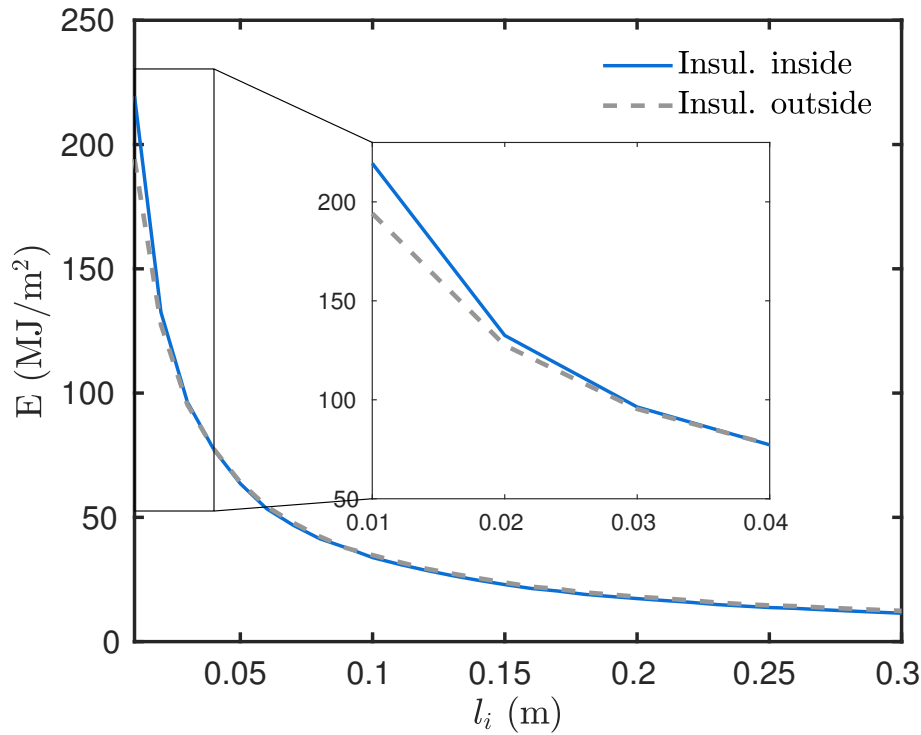}
\caption{\small\em Annual heating and cooling transmission loads for Wall-$2$ and Wall-$3$ in function of insulation thicknesses.}
\label{Figure:parametric_load}
\end{figure}

Simulations are performed with the same numerical and physical configuration of the previous case studies. The annual heating/cooling transmission loads are calculated and presented as a function of $l_{\,i}$ in Figure~\ref{Figure:parametric_load}. As expected the thicker the insulation, the lower the energy consumption for either heating or cooling. One can easily notice that transmission loads drop fast for small insulation thickness values. Besides, for thin insulations, the difference between the two configurations is more evident. Although, this difference is minimized for high values of thicknesses.

With those results, another conclusion from this case study is that the difference of configurations with the insulation inside or outside does not cause important changes on annual heating/cooling when the thickness is higher than $2\ \mathsf{cm}$, despite the thermal mass on the internal side provides important effects on a shorter timescale. For this reason, just for simulation purposes, in what follows, the configuration of Wall$-2$ is used to study the impact of wall-orientation.


\subsection{Third Case: wall orientation influence}
\label{Sec:application_case3}

For the third case study, the same features of the previous case are considered but changing only the facing wall orientation (\textsc{North}, \textsc{South}, \textsc{East}, and \textsc{West}). As \textsc{Brazil} is located in the \textsc{Southern} hemisphere, the \textsc{North} fa\c{c}ade wall of a building is the one that receives more short-wave radiation.

\begin{figure}
\centering
\includegraphics[width=0.69\textwidth]{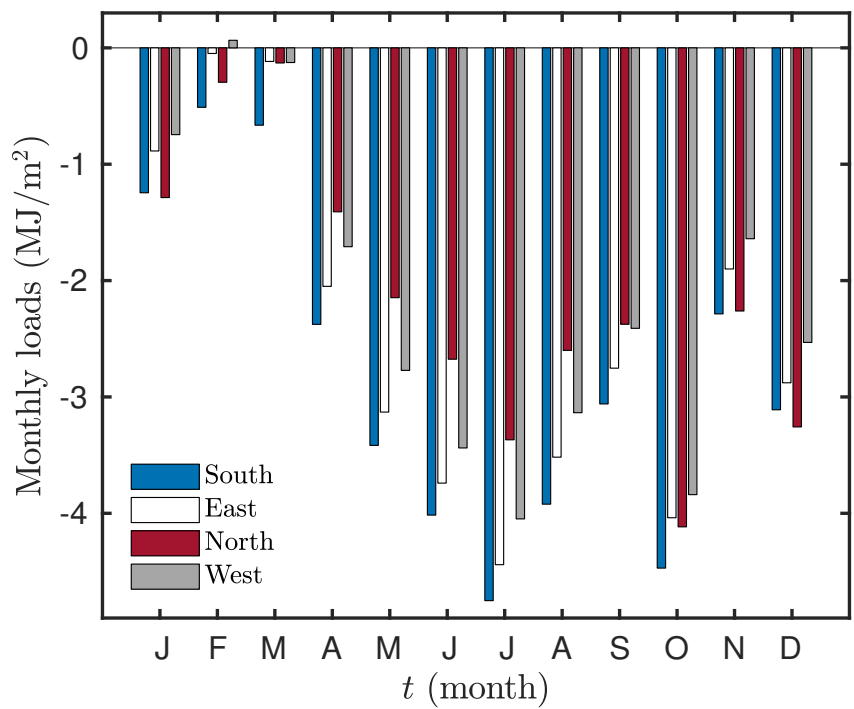}
\caption{\small\em Monthly heating and cooling transmission loads for \textsc{Curitiba} at different orientations.}
\label{Figure:load_wall_NSEW}
\end{figure}

Figure~\ref{Figure:load_wall_NSEW} shows the effect of wall orientation on the monthly heating/cooling transmission loads of Wall$-2$ model with a $10\ \mathsf{cm}$ thick insulation. The monthly transmission loads values are mostly negative indicating a heating demand, which is compatible with the weather of \textsc{Curitiba}. In colder months, the \textsc{South} fa\c{c}ade has the highest heating transmission loads because of the low solar heat gain, which is the opposite for the \textsc{North}-facing wall.

\begin{figure}
\centering
\includegraphics[width=0.69\textwidth]{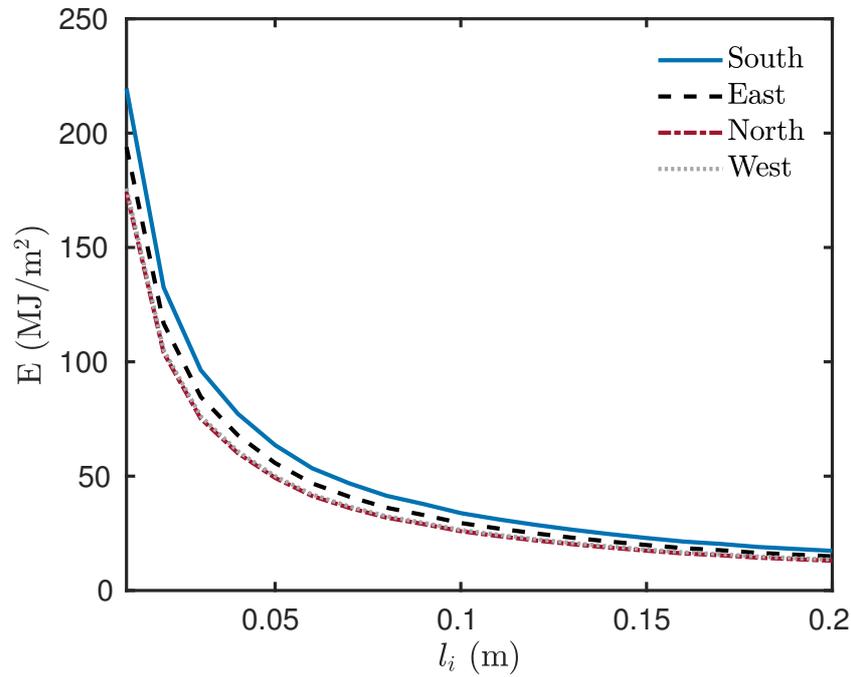}
\caption{\small\em Effect of wall orientation on annual cooling and heating loads in \textsc{Curitiba} for different insulation thickness.}
\label{Figure:Energy_NSEW}
\end{figure}

The effect of wall orientation on yearly heating and cooling transmission loads are presented in function of the insulation thickness in Figure~\ref{Figure:Energy_NSEW}. The \textsc{South}-facing wall has the highest transmission loads, then, \textsc{East}-, \textsc{North}- and \textsc{West}-facing walls with very similar values. As previously observed, transmission loads quickly drop is faster for small values of insulation thickness.

\subsubsection{The roof}

The roof is one of the building components whereby occurs the most significant heat gain/loss. An identical analysis of the wall structure is performed for a roofing system. For this analysis, a flat roof is taken into account and we consider configurations identical to the wall structure, with internal and external insulation. The roofing is made of a $15\ \mathsf{cm}$ concrete fixed layer while the thickness of the insulation material varies. The thermal properties of the materials are reported in Table~\ref{Table:properties}.

Simulations were performed by considering radiative and convective heat transfer at the external roof surface, which exchanges heat with the air and the sky, and receives solar radiation. The solar absorptivity of the outside surface of the roof is taken as $0.5\,$. For the long-wave radiation exchange, a thermal emissivity of $0.9$ is considered and the sky temperature is provided by the software \texttt{Domus} \cite{Mendes2008}. Instead of computing the transmissions loads only for \textsc{Curitiba}, the four cities were comprised in this study.

\begin{figure}
\centering
\includegraphics[width=0.75\textwidth]{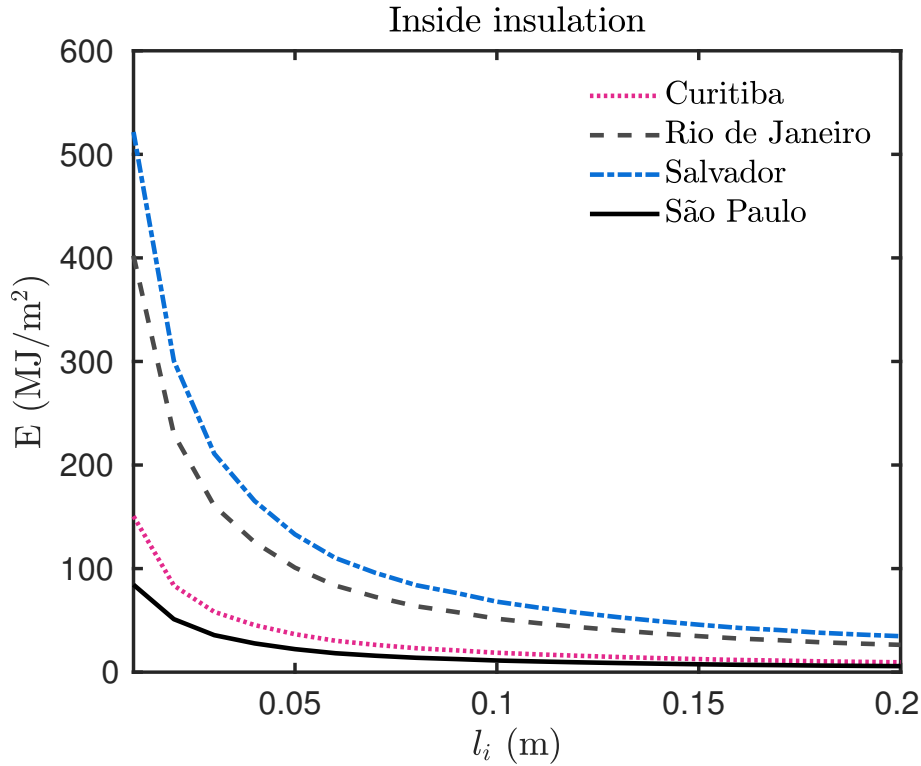}
\caption{\small\em Annual heating/cooling loads associated to the roof for different insulation thicknesses.}
\label{Figure:load_roof}
\end{figure}

Figure~\ref{Figure:load_roof} presents the annual heating/cooling transmission loads as a function of insulation thickness, for the four \textsc{Brazilian} cities. \textsc{Salvador} and \textsc{Rio de Janeiro} are the cities that have the highest transmission loads, which means that they will need a thicker insulation. The differences for the transmission loads are higher for values of insulation thickness. It can also be noticed that the differences are reduced, while the insulation thickness increases.

\subsection{Optimal thermal insulation: economic analysis}

To assess the economic viability of the insulation, a simple economic analysis based on \cite{Wang2000a, Berger2017d} is performed. The total cost of the insulation $C_{\,T}$ is estimated as:
\begin{align*}
  C_{\,T} \egal C_{\,E} \plus C_{\,I} \quad [\mathsf{\$/m^2}]\,,
\end{align*}
where $C_{\,E}$ is the annual cost of the electric energy consumed and $C_{\,I}$ is the the cost of the insulation. The energy cost is calculated by using the annual conduction load $E\,$, the energy efficiency $\eta$ and the energy price $p_{\,e}\,$:
\begin{align*}
  C_{\,E} &\egal  \dfrac{E \cdot p_e}{\eta}  \,.
\end{align*}
The insulation cost $p_{\,i}$ comprises the price of the material as well as its installation and it varies according to the thickness $l_{\,i}\,$: 
\begin{align*}
C_{\,I} &\egal  p_{\,i} \cdot l_{\,i}\,.
\end{align*}

The optimal insulation thickness is given when $C_{\,T}$ has minimum owning and operating costs. For this analysis, the parameters used are given in Table~\ref{Table:economic_parameters}. The energy price is the average price of the electric power in \textsc{Brazil} for $2017$ \cite{ANEEL2017}.

\begin{table}
\centering
\caption{\small\em Economic parameters.}
\bigskip
\setlength{\extrarowheight}{.3em}
\begin{tabular}{ccc}
\hline
Insulation price & Heating/cooling system efficiency & Energy price \\
\hline
$p_i$ & $\eta$ & $p_e$  \\ 
$100\ \mathsf{\$/m^3}$ & $0.8$ & $0.218\ \mathsf{\$/(kWh)}$  \\ 
\hline
\end{tabular}
\label{Table:economic_parameters}
\end{table}

\begin{figure}
\centering
\includegraphics[width=0.75\textwidth]{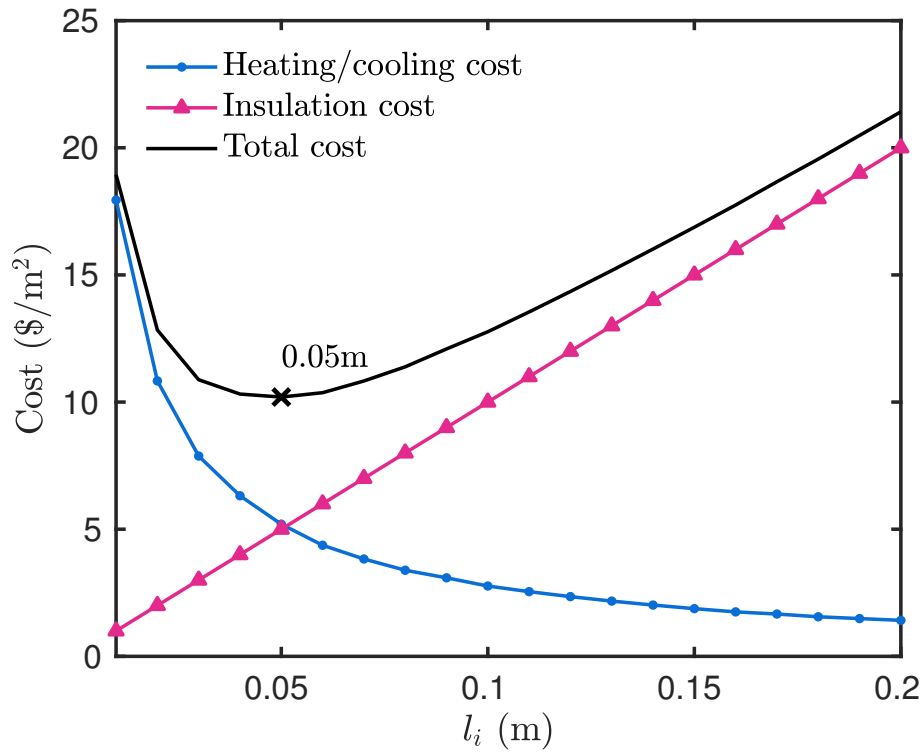}
\caption{\small\em Insulation cost, energy cost and total cost on function of insulation thickness for \textsc{Curitiba} city in a \textsc{South}-facing wall.}
\label{Figure:parametric_oit}
\end{figure}

Figure~\ref{Figure:parametric_oit} presents the insulation cost, the energy consumption cost, and the total cost, as a function of the insulation thickness. As the insulation thickness increases, the energy cost decreases because less energy is used to maintain the temperature inside of the building. However, installation cost increases linearly with the insulation thickness. Thus, the compromise between the energy cost and the insulation cost is the minimum value of their sum. For the case study of Section~\ref{Sec:application_case2_parametric}, it corresponds to an optimum insulation thickness of $0.5\ \mathsf{m}$ for both wall configurations (Wall$-2$ and Wall$-3$).

This economy study is extended to other three cities, considering the four oriented fa\c{c}ades and with the insulation on the inside part of the building wall. Table~\ref{Table:OTI_wall} summarizes the values of optimum insulation thickness for \textsc{East}, \textsc{West}, \textsc{South}, and \textsc{North} oriented walls for the four \textsc{Brazilian} cities. In the case of the wall, the thickness of the insulation is the same for the external and internal configurations. Results show that \textsc{Salvador} is the city that needs the largest insulation thickness, as the inside temperature has the largest difference regarding the outside temperature. \textsc{Salvador} is the hottest city among them with cooling demand along the whole year.

\begin{table}
\centering
\caption{\small\em Optimum insulation thickness according to wall orientation for inside (or outside) insulation.}
\bigskip
\setlength{\extrarowheight}{.3em}
\begin{tabular}{ccccc}
\hline
\textit{Fa\c{c}ade orientation} & \textsc{Curitiba} & \textsc{Rio de Janeiro}  & \textsc{Salvador} & \textsc{S\~ao Paulo} \\
\hline
\textsc{South} & $5\ \mathsf{cm}$ &  $4\ \mathsf{cm}$ & $5\ \mathsf{cm}$ & $3\ \mathsf{cm}$ \\
\textsc{North} & $4\ \mathsf{cm}$ &  $5\ \mathsf{cm}$ & $6\ \mathsf{cm}$ & $2\ \mathsf{cm}$ \\
\textsc{West}  & $4\ \mathsf{cm}$ &  $5\ \mathsf{cm}$ & $6\ \mathsf{cm}$ & $2\ \mathsf{cm}$ \\
\textsc{East}  & $4\ \mathsf{cm}$ &  $4\ \mathsf{cm}$ & $6\ \mathsf{cm}$ & $3\ \mathsf{cm}$ \\
\hline
\end{tabular}
\label{Table:OTI_wall}
\end{table}

The \textsc{Northern} and \textsc{Western} fa\c{c}ades of \textsc{Rio de Janeiro} and \textsc{S\~ao Paulo} are the ones receiving the most solar radiation. In \textsc{Rio de Janeiro}, due to the cooling demand, these fa\c{c}ades require larger insulation thickness, as reported in Table~\ref{Table:OTI_wall}. However, in \textsc{S\~ao Paulo} the heating need prevails, and both fa\c{c}ades orientations have smaller thickness values since their transmission loads are lower. On the other hand, in \textsc{Curitiba} and \textsc{Salvador}, the heat gains are higher for the \textsc{North}, \textsc{West} and \textsc{East} fa\c{c}ades, making the \textsc{South}-facing wall require thicker insulation in \textsc{Curitiba} and thinner in \textsc{Salvador}.

For the roof, the position of the insulation generate differences in the optimum thickness. As shown in Table~\ref{Table:OTI_roof}, \textsc{Curitiba} and \textsc{S\~ao Paulo} have different values of the optimal thickness of external and internal roof insulation. For \textsc{Rio de Janeiro} and \textsc{Salvador}, the thickness of the insulation must be higher, as they exchange more heat through the roof. For all cities, the roof needs more insulation than the wall, as previously mentioned.

\begin{table}
\centering
\caption{\small\em Optimum insulation thickness according to roof insulation position.}
\bigskip
\setlength{\extrarowheight}{.3em}
\begin{tabular}{ccccc}
\hline
\textit{Insulation position} & \textsc{Curitiba} & \textsc{Rio de Janeiro}  & \textsc{Salvador} & \textsc{S\~ao Paulo} \\
\hline
Outside & $3\ \mathsf{cm}$ &  $6\ \mathsf{cm}$ & $7\ \mathsf{cm}$ & $2\ \mathsf{cm}$ \\
Inside & $4\ \mathsf{cm}$ &  $6\ \mathsf{cm}$ & $7\ \mathsf{cm}$ & $3\ \mathsf{cm}$ \\
\hline
\end{tabular}
\label{Table:OTI_roof}
\end{table}


\section{Conclusion}
\label{sec:conclusion}

In this work, the Quasi-Uniform Nonlinear Transformation (QUNT) method has been presented as an innovative method to perform the transient heat transfer, which was applied to an energy optimization problem. The method is based on a non-uniform adaptive grid technique that identifies were the spatial nodes must be placed. It was first compared with a reference solution to verify its features. As a result, a satisfactory accuracy is shown by using much less spatial nodes than traditional methods, by moving its nodes where the temperature gradients are higher \ie, an average ratio of $100$ compared with a uniform grid approach. Another remarkable point of this method is that it is easy to be implemented, by adding extra differential equations without being computational costly. An expressive gain of $25\%$ is obtained.

The high accuracy of the solution instigated the analysis of the physical behavior. The optimum insulation thickness of buildings walls is determined by taking into account the wall orientation and the position of the insulation in \textsc{Brazilian} buildings, which is determined through a parametric study. The annual heating/cooling transmission loads have been calculated using weather data and transient heat flow, simulated over one whole year.

First, a case was focused on comparing a wall with insulation and another one with no insulation, applied to the climatic conditions of one \textsc{Brazilian} city. This case proved the importance of the insulation in reducing the transmission loads. Then, a second case study is simulated to verify if the insulation is better inside or outside of the building wall. The last case considered analyzing the effect of wall orientation and the roof assembly. As \textsc{Brazil} is mostly located in the \textsc{Southern} Hemisphere, the \textsc{North} fa\c{c}ade is the one receiving more solar radiation. In consequence, for the hottest climate cities, the \textsc{North}-facing wall is the one that requires the thicker insulation layers, while, in the \textsc{South} region -- \textsc{Curitiba} --, the \textsc{North}-facing wall requires thinner layers of insulation.

A basic economic analysis was presented, demonstrating that the implementation of an insulation layer appears to be a cost-effective measure to save energy consumption. The optimum insulation thickness for all the wall and roof configurations studied is to be between $2\ \mathsf{cm}$ and $7\ \mathsf{cm}$. The city that requires less insulation is \textsc{S\~ao Paulo}, due to its mild climate, and the city that needs more insulation is \textsc{Salvador}, due to its temperature to be high above the thermal comfort values.

As the proposed numerical method looked very promising, further work is recommended to integrate it to whole-building simulation tools so that more realistic cases, including internal gains and complex boundary conditions, can be taken into account in order to provide more conclusive results and enhance building energy simulation tools.


\section*{Acknowledgements}

This study was financed in part by the Coordena\c{c}\~ao de Aperfei\c{c}oamento de Pessoal de N\'ivel Superior -- Brasil (CAPES) -- in the framework of the International Cooperation Program CAPES/COFECUB (Grant $\#774/13$). The Authors also acknowledge the support from CNRS/INSIS and Cellule \'Energie under the grant MN4BAT--2017. Professor~\textsc{Mendes} thanks the Laboratory LAMA UMR 5127 for the warm hospitality during his visits in 2018, which were supported by the project MN4BAT under the AAP Recherche 2018 programme of the University \textsc{Savoie Mont Blanc}. Finally, the authors acknowledge the Junior Chair Research program ``Building performance assessment, evaluation and enhancement'' from the University \textsc{Savoie Mont Blanc} in collaboration with The French Atomic and Alternative Energy Center (CEA) and Scientific and Technical Center for Buildings (CSTB).


\newpage
\section*{Nomenclature}

\begin{table}[h!]
\centering
\begin{tabular*}{0.7\textwidth}{@{\extracolsep{\fill}} |@{} >{\scriptsize} c >{\scriptsize} l >{\scriptsize} l| }
\hline 
\multicolumn{3}{|c|}{\textbf{Nomenclature}}\\[3pt]
& \multicolumn{2}{l|}{\emph{Latin letters}}\\[3pt]
$\cz$ & material specific heat capacity & $[\mathsf{J/(kg\cdot K)}]$ \\
$c$ & energy storage coefficient & $[\mathsf{J/(m^3\cdot K)}]$ \\
$E$ & heat transmission load & $[\mathsf{MJ/m^2}]$ \\
$h$ & convective heat transfer coefficient & $[\mathsf{W/(m^2\cdot K)}]$ \\
$k$ & thermal conductivity & $[\mathsf{W/(m\cdot K)}]$ \\
$l$ & wall length & $[\mathsf{m}]$ \\
$q$ & heat flux & $[\mathsf{W/m^2}]$ \\
$T$ & temperature & $[\mathsf{K}]$ \\
$t$ & time & $[\mathsf{s}]$ \\
$x$ & thickness coordinate direction & $[\mathsf{m}]$ \\
& \multicolumn{2}{l|}{\emph{Greek letters}} \\[3pt]
$\alpha$ & solar absorptivity of outside surface wall & $[-]$ \\
$\epsilon$ & emissivity & $[-]$\\
$\rho$ & material density & $[\mathsf{kg/m^3}]$ \\
$\sigma$ & \textsc{Stefan-Boltzmann} constant & $[\mathsf{W/(m^2\cdot K^4)}]$\\
& \multicolumn{2}{l|}{\emph{Dimensionless parameters}}\\[3pt]
$\mathrm{Bi}$ & \textsc{Biot} number & $[-]$ \\
$\cTs$ & storage coefficient & $[-]$ \\
$\mathrm{Fo}$ & \textsc{Fourier} number & $[-]$ \\
$\kTs$ & diffusion coefficient & $[-]$ \\
$\qsinf$ & short-wave radiation source term  & $[-]$ \\
$\Rlws$ & long-wave radiation coefficient & $[-]$ \\
$u$ & temperature field & $[-]$ \\
\hline
\end{tabular*}
\end{table}

\bigskip\bigskip
\addcontentsline{toc}{section}{References}
\bibliographystyle{abbrv}
\bibliography{biblio}
\bigskip\bigskip

\end{document}